\documentstyle[amstex,amssymb,12pt,epsf]{article}

\begin{document}

\begin{center}
{\bf {\Large TIME-REVERSAL-VIOLATING OPTICAL GYROTROPY}}

\bigskip

Vladimir G.Baryshevsky

Nuclear Problem Research Institute,

Bobruiskaya Str.11, Minsk 220080 Belarus.

Electronic address: bar@@inp.minsk.by

Tel: 00375-172-208481, Fax: 00375-172-265124

\bigskip
\end{center}

Time-reversal-violating interactions of the electrons and nucleus cause the
appearance of new optical phenomena. These phenomena are not only very
interesting from fundamental point of view, but give us a new key for
studying the time-reversal-violating interactions of the elementary
particles.

Violation of time reversal symmetry has been observed only in $K_{0}$-decay
many years ago \cite{1}, and remains one of the great unsolved problems in
elementary particle physics. Since the discovery of the CP-violation in
decay of $K_{0}$-mesons, a few attempts have been undertaken to observe this
phenomenon experimentally in different processes. However, those experiments
have not been successful. At the present time novel more precise
experimental schemes are actively discussed: observation of the atom \cite{2}
and neutron \cite{3} electric dipole moment (EDM) through either spin
precession or light polarization plane rotation caused by pseudo-Zeeman
splitting of atom (molecule) levels by external electric field \ $%
\overrightarrow{E}$\ due to interaction of atom level \ EDM $\overrightarrow{%
d_{a}}$\ with electric field\ $W=-\overrightarrow{d_{a}}\cdot 
\overrightarrow{E}$ \cite{Budker,B3,B17,Sushkov,Zeldovich}(this effect is
similar to magneto-optic effect Macaluso-Corbino \cite{Macaluso}). It should
be noted that the mentioned experiments use the possible existence of such
intrinsic quantum characteristic of atom (molecule) as \ static EDM.
According to \cite{4,5,6} together with the EDM there is one more
characteristic of atom (molecule) describing its response to the external
field effect - the T- and P-odd polarizability of atom (molecule) $\beta
^{T} $. This polarizability differs from zero even if EDM of electron is
equal to zero and pseudo-Zeeman splitting of atom (molecule) levels is
absent.{\em \ }According to \cite{6, lanl-last,PLA} the T-odd polarizability 
$\beta ^{T}$ yields to the appearance of new optical phenomena - photon
polarization plane rotation and circular dichroism in an optically
homogeneous isotropic medium exposed to an electric field caused by the
Stark mixing of atom (molecule) levels. This T-odd phenomenon is a kinematic
analog of the well known T-even phenomenon of \ Faraday effect of the photon
polarization plane rotation in the medium exposed to a magnetic field due to
Van-Vleck mechanism. Similarly \ to the well known P-odd T-even \ effect of
light polarization plane rotation for which the intrinsic spin spiral of
atom is responsible \cite{7}, this effect is caused by the atom
magnetization appearing under external electric field action (see section 3
below). \ Moreover, according to \cite{lanl99} and section 3 below, the
magnetization of atom appearing under action of static electric field causes
the appearance of induced magnetic field \ $\overrightarrow{H_{ind}}$ . The
energy of interaction of \ atom magnetic moment $\overrightarrow{\mu _{a}}$\
with this field is \ $W_{H}=-\overrightarrow{\mu _{a}}\cdot $\ $%
\overrightarrow{H_{ind}}(\overrightarrow{E}).$\ Therefore, the total
splitting of atom levels is determined by energy {\em \ }$W_{T}=-%
\overrightarrow{d_{a}}\cdot \overrightarrow{E}-\overrightarrow{\mu _{a}}%
\cdot \overrightarrow{H_{ind}}(\overrightarrow{E})$. As a result, the effect
of polarization plane rotation deal with the energy levels splitting is
caused not only by $\overrightarrow{d_{a}}$ interaction with electric field\
but by $\overrightarrow{H_{ind}}(\overrightarrow{E})$ interaction with $%
\overrightarrow{\mu _{a}}$, too.\ It is easy to see, that even for $%
\overrightarrow{d_{a}}=0$\ the energy of splitting differs from zero and the
T-odd effect of polarization plane rotation exists. One more interesting
T-odd phenomenon appears when the photon beam is incident orthogonally to
the external electric field \ $\overrightarrow{E}$\ \ (or magnetic field \ $%
\overrightarrow{H}$ or both electric and magnetic fields). This is
birefringence effect \cite{lanl-last} (i.e. the effect when plane polarized
photons are converted to circular polarized ones and vice versa).

Also the T-odd phenomenon of photon polarization plane rotation and circular
\ dichroism appears at photon passing through non-center-symmetrical
diffraction grating \cite{6}.

\begin{figure}[hbp]
\epsfxsize = 13 cm \centerline{\epsfbox{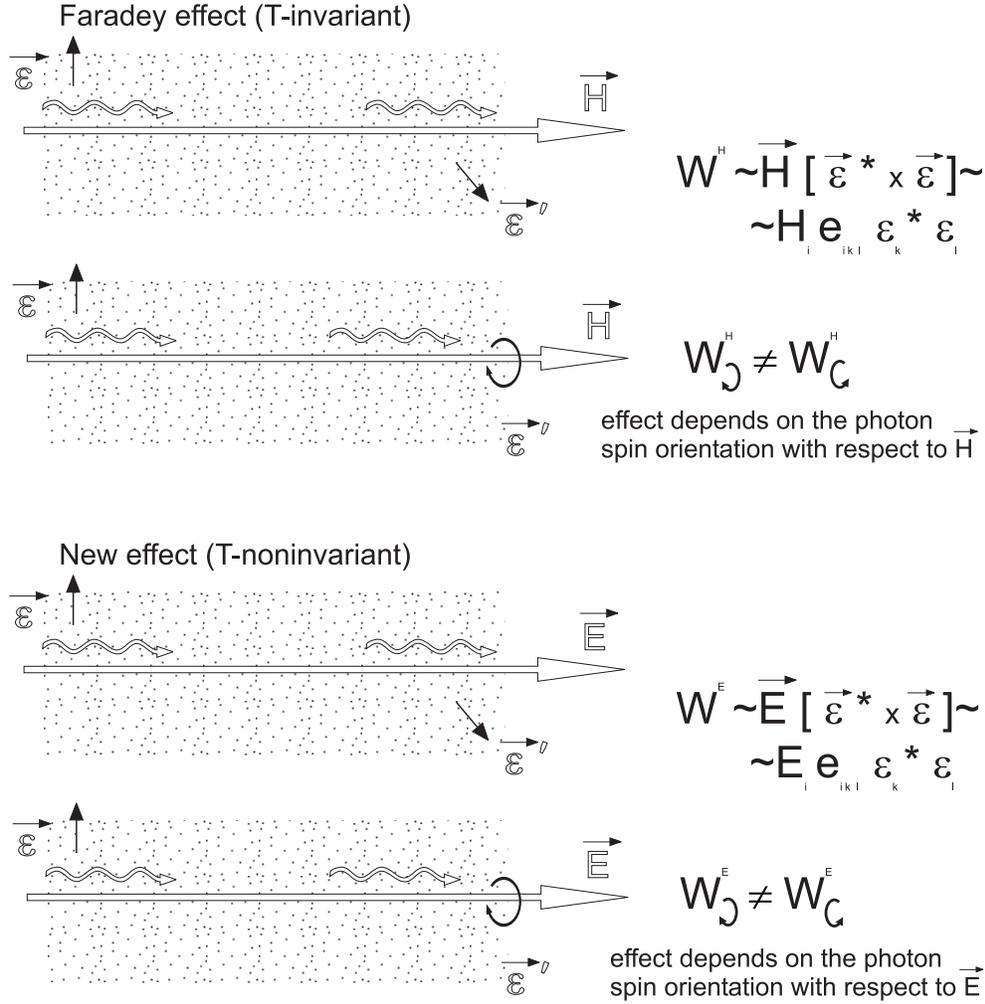}}
\caption{Phenomenon of the time-reversal-violating photon polarization plane
rotation and circular dichroism by an electric field.}
\label{faradey}
\end{figure}

\begin{figure}[htbp]
\epsfxsize = 10 cm \centerline{\epsfbox{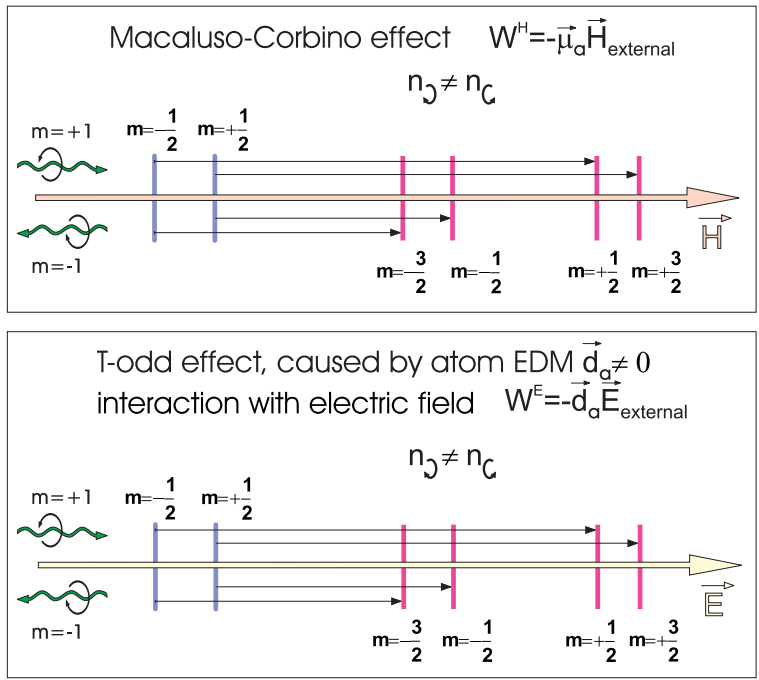}}
\caption{Polarization plane rotation phenomena.}
\label{macaluso}
\end{figure}

\begin{figure}[htbp]
\epsfxsize = 10 cm \centerline{\epsfbox{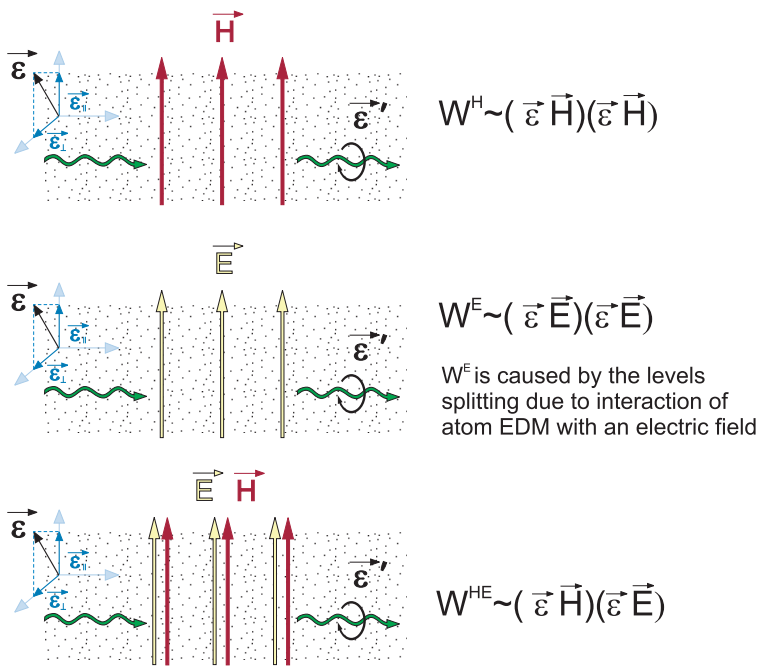}}
\caption{T-noninvariant birefringence effect.}
\label{birefringence}
\end{figure}

\begin{figure}[htp]
\epsfxsize = 14 cm \centerline{\epsfbox{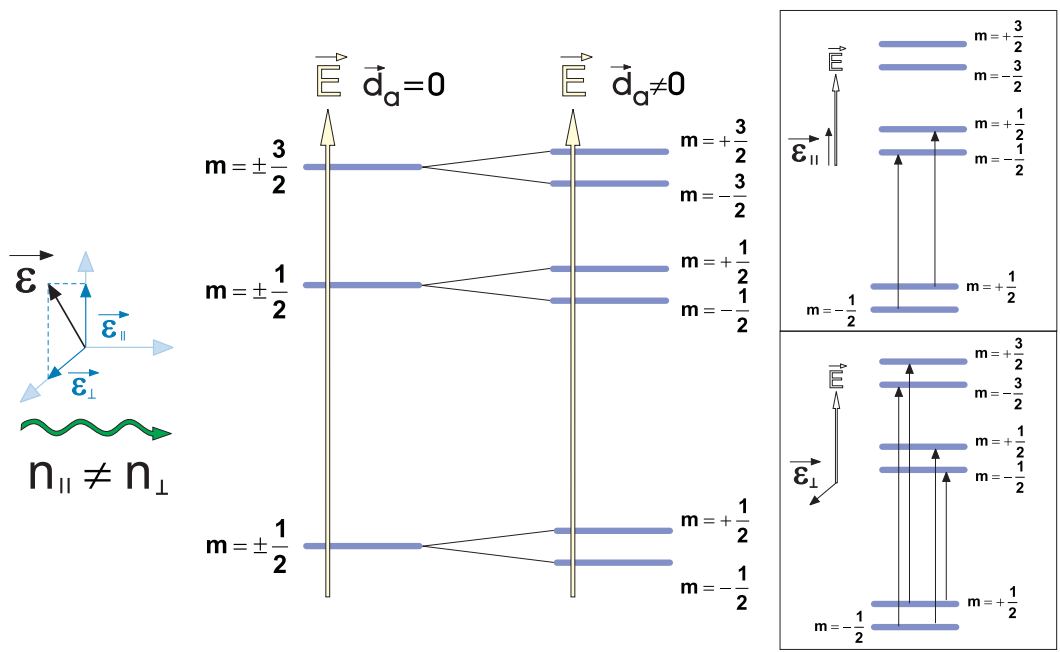}}
\caption{Levels splitting.}
\label{levels}
\end{figure}

\clearpage
\newpage

{\bf {\Large 1. Phenomenon of the time-reversal-violating photon
polarization plane rotation by an electric field.}}

\bigskip

In this section the T-odd phenomenon of the photon polarization plane
rotation (circular dichroism) in an atomic (molecular) gas exposed to an
electric field is considered. The expression for the T-non-invariant
polarizability of an atom (molecule) in an electric field is obtained. It is
shown that the T-odd rotation angle $\vartheta ^{T}$ grows up when the
energy of interaction of an atom (molecule) with an electric field is of the
order of the opposite parity levels spacing.

In accordance with \cite{4,5,6}\ the T-reversal violating dielectric
permittivity tensor $\varepsilon _{ik}$\ for an optically diluted medium ( $%
\varepsilon _{ik}-\delta _{ik}\ll 1$, \ $\delta _{ik}$\ is the Kronecker
symbol) is given by
\begin{equation}
\varepsilon _{ik}=\delta _{ik}+\chi _{ik}=\delta _{ik}+\frac{4\pi \rho }{%
k^{2}}f_{ik}(0),
\end{equation}
where $\chi _{ik}$\ is the polarizability tensor of a medium, $\rho $\ is
the number of atoms (molecules) per $cm^{3}$, $k$\ is the photon wave
number; $f_{ik}(0)$\ is the tensor part of the zero angle amplitude of
elastic coherent scattering of a photon by an atom (molecule) $%
f(0)=f_{ik}(0)e_{i}^{\prime \ast }e_{k}$. Here $\overrightarrow{e}$\ and $%
\overrightarrow{e}^{\prime }$ \ are the polarization vectors of initial and
scattered photons. Indices $i=1,2,3$ are referred to coordinates $x,y,z$,
respectively, repeated indices imply summation.

Let photon be scattered by nonpolarized atoms (molecules) interacting with
an electric field $\overrightarrow{E}$. When photon propagates along the
electric field the amplitude $f_{ik}(0)$\ can be written as \cite{6}:
\begin{equation}
f_{ik}(0)=f_{ik}^{ev}+\frac{\omega ^{2}}{c^{2}}[i\beta _{s}^{P}\varepsilon
_{ikl}n_{\gamma l}+i\beta _{E}^{T}\varepsilon _{ikl}n_{El}],
\end{equation}
where $f_{ik}^{ev}$\ is the T-, P-even (invariant) part of \ $f_{ik}(0)$\ , $%
\beta _{s}^{P}$\ is the scalar P-violating polarizability of an atom
(molecule), $\beta _{E}^{T}$\ is the scalar T- and P-violating
polarizability of an atom (molecule), $\varepsilon _{ikl}$\ is the totally
antisymmetric unit tensor of the rank three, $\overrightarrow{n}_{\gamma }=%
\frac{\overrightarrow{k}}{k}$\ , $\overrightarrow{k}$\ is the photon wave
vector, $\overrightarrow{n}_{E}=\frac{\overrightarrow{E}}{E}$\ .

The term proportional to $\beta _{s}^{P}$ describes the T-invariant
P-violating photon polarization plane rotation (and circular dichroism) in a
gas \cite{7}. The corresponding refractive index $N$\ is as follows:
\begin{equation}
N=1+\frac{2\pi \rho }{k^{2}}\left\{ f_{ik}^{ev}(0)e_{i}^{\ast }e_{k}+i\frac{%
\omega ^{2}}{c^{2}}(\beta _{s}^{P}\overrightarrow{n}_{\gamma }+\beta _{E}^{T}%
\overrightarrow{n}_{E})[\overrightarrow{e}^{\ast }\times \overrightarrow{e}%
]\right\} .  \label{3}
\end{equation}
The unit vectors describing the circular polarization of photons are: $%
\overrightarrow{e}_{+}=-\frac{\overrightarrow{e}_{1}+i\overrightarrow{e}_{2}%
}{\sqrt{2}}\ $for the right and, $\overrightarrow{e}_{-}=\frac{%
\overrightarrow{e}_{1}-i\overrightarrow{e}_{2}}{\sqrt{2}}$\ for the left
circular polarization, where $\overrightarrow{e}_{1}\perp \overrightarrow{e}%
_{2}\ $, $\overrightarrow{e}_{2}=[\overrightarrow{n}_{\gamma }\times 
\overrightarrow{e}_{1}]\ $are the unit polarization vectors of a linearly
polarized photon, $[\overrightarrow{e}_{1}\times \overrightarrow{e}_{2}]=%
\overrightarrow{n}_{\gamma }$, $\overrightarrow{e}_{1}=-\frac{%
\overrightarrow{e}_{+}-\overrightarrow{e}_{-}}{\sqrt{2}}$, $\overrightarrow{e%
}_{2}=-\frac{\overrightarrow{e}_{+}+\overrightarrow{e}_{-}}{i\sqrt{2}}$.

Let an electromagnetic wave propagates through a gas along the electric
field\ $\overrightarrow{E}$ direction. The refractive indices for the right $%
N_{+}$\ and for the left $N_{-}$\ circular polarized photons can be written
as:
\begin{equation}
N_{\pm }=1+\frac{2\pi \rho }{k^{2}}f_{\pm }(0)=1+\frac{2\pi \rho }{k^{2}}%
\left\{ f^{ev}(0)\mp \frac{\omega ^{2}}{c^{2}}\left[ \beta _{s}^{P}+\beta
_{E}^{T}(\overrightarrow{n}_{E}\overrightarrow{n}_{\gamma })\right] \right\}
,
\end{equation}
where $f_{+}(0)(f_{-}(0))$ is the zero angle amplitude of the elastic
coherent scattering of the right (left) circular polarized photon by an atom
(molecule).

Let photons with the linear polarization $\overrightarrow{e}_{1}=-\frac{%
\overrightarrow{e}_{+}-\overrightarrow{e_{-}}}{\sqrt{2}}$\ fall in a gas.
The polarization vector of a photon in a gas $\overrightarrow{e}_{1}^{\prime
}$\ can be written as: 
\begin{eqnarray}
\overrightarrow{e}_{1}^{\prime }\ &=&-\frac{\overrightarrow{e}_{+}\ }{\sqrt{2%
}}e^{ikN_{+}L}+\frac{\overrightarrow{e}_{-}\ }{\sqrt{2}}e^{ikN_{-}L}=
\label{5} \\
&=&e^{\frac{1}{2}ik(N_{+}+N_{-})L}\left\{ \overrightarrow{e}_{1}\cos \frac{1%
}{2}k(N_{+}-N_{-})L-\overrightarrow{e}_{2}\sin \frac{1}{2}%
k(N_{+}-N_{-})L\right\} ,  \nonumber
\end{eqnarray}
where $L$\ is the photon propagation length in a medium.

As one can see, the photon polarization plane rotates in a gas. The angle of
rotation $\vartheta $\ is 
\begin{eqnarray}
\vartheta &=&\frac{1}{2}k {Re}(N_{+}-N_{-})L=\frac{\pi \rho }{k} {Re}\left[
f_{+}(0)-f_{-}(0)\right] L=  \label{6} \\
&=&-\frac{2\pi \rho \omega }{c}\left[ \beta _{s}^{P}+\beta _{E}^{T}(%
\overrightarrow{n}_{E}\overrightarrow{n}_{\gamma })\right] L,  \nonumber
\end{eqnarray}
where ${Re}N_{\pm }$\ is the real part of $N_{\pm }$. It should be noted
that $\vartheta >0$\ corresponds to the right rotation of the light
polarization plane and $\vartheta <0$\ corresponds to the left one, where
the right (positive) rotation is recording by the light observer as the
clockwise one.

In accordance with (\ref{6}) the T-odd interaction results in the photon
polarization plane rotation around the electric field $\overrightarrow{E}$%
direction. The angle of rotation is proportional to the
polarizability $\beta _{E}^{T}$ and the $(\overrightarrow{n}_{E}%
\overrightarrow{n}_{\gamma })$ correlation. Together with the T-odd effect
there is \ the T-even P-odd polarization plane rotation phenomenon
determining by the polarizability $\beta _{s}^{P}$ and being independent on
the $(\overrightarrow{n}_{E}\overrightarrow{n}_{\gamma })$ correlation. The
T-odd rotation dependence on the electric field {\LARGE \ }$\overrightarrow{E%
}$ orientation with respect to the $\overrightarrow{n}_{\gamma }$ direction
allows one to distinguish T-odd and T-even P-odd phenomena experimentally.

The refractive index $N_{+}(N_{-})$\ has both real and imaginary parts. The
imaginary part of the refractive index $({Im}N_{\pm }\sim {Im}\beta _{E}^{T}(%
\overrightarrow{n}_{E}\overrightarrow{n}_{\gamma }))$\ is responsible for
the T-reversal violating circular dichroism. Due to this process the
linearly polarized photon takes circular polarization. The sign of the
circular polarization depends on the sign of the scalar production $(%
\overrightarrow{n}_{E}\overrightarrow{n}_{\gamma })$\ that allows us to
separate T-odd circular dichroism from P-odd T-even circular dichroism. The
last one is proportional to ${Im}\beta _{s}^{P}$.

Before the detailed description of the effect let us consider
simple observations to demonstrate clearly how the effect appears (see also \cite{Mats} ). Let us
expose atom with a single valence electron being in a ground state $S_{1/2}$%
\ to an electric field. $P$\ and $T$\ odd interactions and interaction with
electric field yield to the admixture of opposite parity states to the
ground state. Considering only mixing with the nearest $nP_{1/2}$\ state one
can write atom wave function $|\widetilde{s}_{1/2}>$ 
\begin{equation}
|{\widetilde{s}_{1/2}}>=\frac{1}{\sqrt{4\pi }}(R_{0}(r)-R_{1}(r)(\vec{\sigma}%
\vec{n})\eta -R_{1}(r)(\vec{\sigma}\vec{n})(\vec{\sigma}\vec{E})\delta
)|\chi _{1/2}\rangle ,
\end{equation}
here $\vec{\sigma}$\ - are the Pauli matrices, $\vec{n}=\vec{r}/r$\ is the
unit vector along $\vec{r}$ direction, $R_{0}$\ and $R_{1}$\ are the radial
parts of $nS_{1/2}$\ and $nP_{1/2}$\ wave functions, respectively, $|\chi
_{1/2}\rangle $\ is the spin part of wave function, $\eta $\ is the mixing
coeffiecient for $P_{1/2}$\ state due to $P$\ and $T$\ noninvariant
interactions and $\delta $\ is that caused by electric field.

\begin{figure}[htbp]
\epsfxsize =7 cm \centerline{\epsfbox{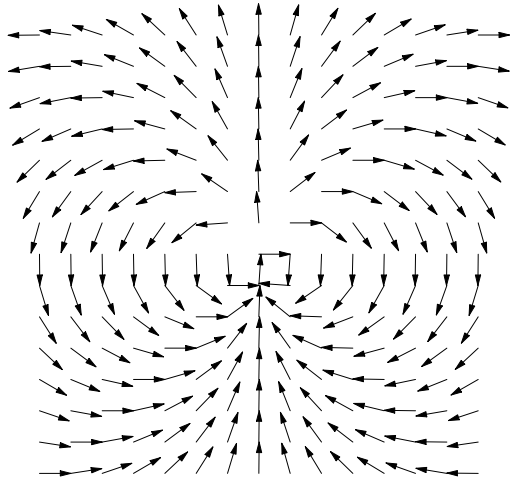}} 
\vspace{0.5cm}
\caption{{Vector field $4\vec{n}(\vec{n}\vec{E})-2\vec{E}$. Vectors in
figure show direction of atom spin in $S_{1/2}$ state with the admixture of 
$P_{1/2}$ state due to $PT$-noninvariant interactions
and external electric field.}}
\label{figb}
\end{figure}

Let us consider orientation of electron spin in atom. In order to find the
spatial distribution of spin direction one should average spin operator over
spin part of atom wave function. Only terms containing both $\delta $ and $%
\eta $ contribute to $PT$-odd rotation of polarization plane of light. The
change of spin direction due to these terms is  
\begin{eqnarray}
\Delta \vec{s}(\vec{r}) &=&\frac{\eta \delta }{8\pi }R_{1}^{2}\left\langle
\chi _{1/2}|(\vec{\sigma}\vec{n})\vec{\sigma}(\vec{\sigma}\vec{n})(\vec{%
\sigma}\vec{E})+(\vec{\sigma}\vec{E})(\vec{\sigma}\vec{n})\vec{\sigma}(\vec{%
\sigma}\vec{n})|\chi _{1/2}\right\rangle   \nonumber \\
&=&\frac{\eta \delta R_{1}^{2}}{8\pi }\left( 4\vec{n}(\vec{n}\vec{E})-2\vec{E%
}\right) 
\end{eqnarray}
The vector field ($4\vec{n}(\vec{n}\vec{E})-2\vec{E})$\ is shown in Fig.\ref{figb}.
Since $\Delta \vec{s}$\ does not depend on initial direction of atom spin,
this spin structure appears even in non-polarized atom. It should be noted
that spin vector averaged over spatial variables differs from zero and is
directed along the electric field $\vec{E}$. Photons with angular moment
parallel and antiparallel to the electric field can interact with such spin
structure in a different ways that causes rotation of polarization plane of
light. 

In order to estimate the magnitude of the effect one should obtain the T-odd
polarizability $\beta _{E}^{T}$\ or (that is actually the same, see (\ref{2},%
\ref{6})) the amplitude $f_{\pm }(0)$ of elastic coherent scattering of a
photon by an atom (molecule). According to quantum electrodynamics the
elastic coherent scattering at zero angle can be considered as the
succession of two processes: the first one is the absorption of the initial
photon with the momentum $\overrightarrow{k}$ and the transition of the atom
(molecule) from the initial state $\left| N_{0}\right\rangle $ with the
energy $E_{N_{0}}$ to an intermediate state $\left| F\right\rangle $ with an
energy $E_{F}$; the second one is the transition of the atom (molecule) from
the state $\left| F\right\rangle $ to the final state $\left| F^{\prime
}\right\rangle =\left| N_{0}\right\rangle $ and irradiation of the photon
with the momentum $\ \overrightarrow{k}^{\prime }=\overrightarrow{k}$.

Let $H_{A}$\ be the atom (molecule) Hamiltonian considering the weak
interaction between electrons and nucleus and the electromagnetic
interaction of an atom (molecule) with the external electric $%
\overrightarrow{E}$ and magnetic $\overrightarrow{H}$ fields. It defines the
system of eigenfunctions $\left| F\right\rangle $\ and eigenvalues $%
E_{F}=E_{F}(\overrightarrow{E,}\overrightarrow{H})$: 
\begin{equation}
H_{A}\left| F\right\rangle =E_{F}\left| F\right\rangle ,  \label{7}
\end{equation}

$F-$set of quantum numbers describing the state $\left| F\right\rangle $:

The matrix element of the process determining the scattering amplitude in
the forward direction in the dipole approximation is given by \cite{8}: 
\begin{equation}
{\frak M}_{N_{0}}=\sum_{F}\left\{ \frac{\left\langle N_{0}\right| 
\overrightarrow{d}\overrightarrow{e}^{\ast }\left| F\right\rangle
\left\langle F\right| \overrightarrow{d}\overrightarrow{e}\left|
N_{0}\right\rangle }{E_{F}-E_{N_{0}}-\hbar \omega }+\frac{\left\langle
N_{0}\right| \overrightarrow{d}\overrightarrow{e}\left| F\right\rangle
\left\langle F\right| \overrightarrow{d}\overrightarrow{e}^{\ast }\left|
N_{0}\right\rangle }{E_{F}-E_{N_{0}}+\hbar \omega }\right\} ,  \label{8}
\end{equation}
where $\overrightarrow{d}$\ is the electric dipole transition operator, $%
\omega $\ is the photon frequency.

It should be reminded that the atom (molecule) exited levels are
quasistationary: $E_{F}=E_{F}^{(0)}-\frac{i}{2}\Gamma _{F}$\ , $E_{F}^{(0)}$%
\ is the atom (molecule) level energy, \ $\Gamma _{F}$ is the level width.
The matrix element (\ref{8}) can be written as: 
\begin{equation}
{\frak M}_{N_{0}}=\alpha _{ik}^{N_{0}}e_{i}^{\ast }e_{k},  \label{9}
\end{equation}
where $\alpha _{ik}^{N_{0}}$ is the tensor of dynamical polarizability of an
atom (molecule)
\begin{equation}
\alpha _{ik}^{N_{0}}=\sum_{F}\left\{ \frac{\left\langle N_{0}\right|
d_{i}\left| F\right\rangle \left\langle F\right| d_{k}\left|
N_{0}\right\rangle }{E_{F}-E_{N_{0}}-\hbar \omega }+\frac{\left\langle
N_{0}\right| d_{k}\left| F\right\rangle \left\langle F\right| d_{i}\left|
N_{0}\right\rangle }{E_{F}-E_{N_{0}}+\hbar \omega }\right\}  \label{10}
\end{equation}
In general case atoms are distributed to the levels of ground state $N_{0}$\
with the probability $P(N_{0})$. Therefore, $\alpha _{ik}^{N_{0}}$\ should
be averaged over all states $N_{0}$. As a result, the polarizability can be
written
\begin{equation}
\alpha _{ik}=\sum_{N_{0}}P(N_{0})\alpha _{ik}^{N_{0}}  \label{10-1}
\end{equation}
The tensor $\alpha _{ik}$ can be expanded in the irreducible parts as 
\begin{equation}
\alpha _{ik}=\alpha _{0}\delta _{ik}+\alpha _{ik}^{s}+\alpha _{ik}^{a},
\label{11}
\end{equation}
where $\alpha _{0}=\frac{1}{3}{\sum_{i} \alpha _{ii}}$ is the scalar, $%
\alpha _{ik}^{s}=\frac{1}{2}(\alpha _{ik}+\alpha _{ki})-$\ $\alpha
_{0}\delta _{ik}$\ is the symmetric tensor of rank two, $\alpha _{ik}^{a}=%
\frac{1}{2}(\alpha _{ik}-\alpha _{ki})$ is the antisymmetric tensor of rank
two, 
\begin{equation}
\alpha _{ik}^{a}=\frac{\omega }{\hbar }\sum_{N_{0}}P(N_{0})\sum_{F}\left\{ 
\frac{\left\langle N_{0}\right| d_{i}\left| F\right\rangle \left\langle
F\right| d_{k}\left| N_{0}\right\rangle -\left\langle N_{0}\right|
d_{k}\left| F\right\rangle \left\langle F\right| d_{i}\left|
N_{0}\right\rangle }{\omega _{FN_{0}}^{2}-\omega ^{2}}\right\} \text{,}
\label{12}
\end{equation}
where $\omega _{FN_{0}}=\frac{E_{F}-E_{N_{0}}}{\hbar }$.

Let atoms (molecules) be nonpolarized. The antisymmetric part of
polarizability (\ref{12}) is equal to zero in the absence of $TP$-odd
interactions and external magnetic field. It should be reminded that for the
P-odd and T-even interactions the antisymmetric part of polarizability
differs from zero only for both the electric and magnetic dipole transitions 
\cite{7}.

For further analysis suppose the external magnetic field be equal to zero
(electric field differs from zero). We can evaluate the antisymmetric part $%
\alpha _{ik}^{a}$\ of the tensor $\alpha _{ik}$ of dynamical polarizability
of atom (molecule), and, as a result, obtain the expression for $\beta
_{E}^{T}$\ by the following way. According to (4,6) the magnitude of the
T-odd effect is determined by the polarizability $\beta _{E}^{T}$\ or (that
is actually the same, see (\ref{2},\ref{6})) by the amplitude $f_{\pm }(0)$\
of elastic coherent scattering of a photon by an atom (molecule). If $%
\overrightarrow{n}_{E}\parallel \overrightarrow{n}_{\gamma }$\ the amplitude 
$f_{\pm }(0)$\ in the dipole approximation can be written as $f_{\pm }=\mp 
\frac{\omega ^{2}}{c^{2}}\beta _{E}^{T}$. As a result, in order to obtain
the amplitude $f_{\pm }$, the matrix element (\ref{8},\ref{9}) for photon
polarization states $\overrightarrow{e}=\overrightarrow{e}_{\pm }$\ should
be found.

The electric dipole transition operator$\overrightarrow{d}$ can be written
in the form: 
\begin{equation}
\overrightarrow{d}=d_{+}\overrightarrow{e}_{+}+d_{-}\overrightarrow{e}%
_{-}+d_{z}\overrightarrow{n}_{\gamma }\text{,}  \label{13}
\end{equation}
with $\overrightarrow{d}_{+}=-\frac{d_{x}-id_{y}}{\sqrt{2}}$, $%
\overrightarrow{d}_{-}=\frac{d_{x}+id_{y}}{\sqrt{2}}$. Let photon
polarization state $\overrightarrow{e}=\overrightarrow{e}_{+}$. Using (\ref
{8},\ref{9}) we can present the polarizability $\beta _{E}^{T}$ as follows: 
\begin{equation}
\beta _{E}^{T}=\frac{\omega }{\hbar }\sum_{N_{0}}P(N_{0})\sum_{F}\left\{ 
\frac{\left\langle N_{0}\right| d_{-}\left| F\right\rangle \left\langle
F\right| d_{+}\left| N_{0}\right\rangle -\left\langle N_{0}\right|
d_{+}\left| F\right\rangle \left\langle F\right| d_{-}\left|
N_{0}\right\rangle }{\omega _{FN_{0}}^{2}-\omega ^{2}}\right\} .  \label{14}
\end{equation}
For further analysis the more detailed expressions for atom (molecule) wave
functions are necessary. The weak interaction constant is very small.
Therefore, we can use the perturbation theory. Let $\left| f,E\right\rangle $
be the wave function of an atom (molecule) interacting with an electric
field $\overrightarrow{E}$\ in the absence of weak interaction . Switch on
weak interaction $(V_{w}\neq 0)$. According to the perturbation theory\ \cite
{8}\ the wave function $\left| F\right\rangle $\ can be written in this case
as: 
\begin{equation}
\left| F\right\rangle =\left| f,\overrightarrow{E}\right\rangle +\sum_{n}%
\frac{\left\langle n,\overrightarrow{E}\right| V_{w}\left| f,\overrightarrow{%
E}\right\rangle }{E_{f}-E_{n}}\left| n,\overrightarrow{E}\right\rangle
\label{15}
\end{equation}
It should be mentioned that both numerator and denominator of (\ref{14})
contain $V_{w}$. Suppose $V_{w}$\ to be small one can represent the total
polarizability $\beta _{E}^{T}$\ as the sum of two terms
\begin{equation}
\beta _{E}^{T}=\beta _{mix}^{T}+\beta _{split}^{T},  \label{sum}
\end{equation}
where 
\begin{equation}
\beta _{mix}^{T}=\frac{\omega }{\hbar }\sum_{N_{0}}P(N_{0})\sum_{f}\frac{1}{%
\omega _{fn_{0}}^{2}-\omega ^{2}}\sum_{l}  \label{16}
\end{equation}

\bigskip $\left\{ \frac{2 {Re}\left[ \left\{ \left\langle n_{0}%
\overrightarrow{E}\right| d_{-}\left| f\overrightarrow{E}\right\rangle
\left\langle f\overrightarrow{E}\right| d_{+}\left| l\overrightarrow{E}%
\right\rangle -\left\langle n_{0}\overrightarrow{E}\right| d_{+}\left| f%
\overrightarrow{E}\right\rangle \left\langle f\overrightarrow{E}\right|
d_{-}\left| l\overrightarrow{E}\right\rangle \right\} \left\langle l%
\overrightarrow{E}\right| V_{w}\left| n_{0}\overrightarrow{E}\right\rangle %
\right] }{E_{n_{0}}-E_{l}}\right. +$

\bigskip$\left. +\frac{2 {Re}\left[ \left\langle n_{0}\overrightarrow{E}%
\right| d_{-}\left| l\overrightarrow{E}\right\rangle \left\langle l%
\overrightarrow{E}\right| V_{w}\left| f\overrightarrow{E}\right\rangle
\left\langle f\overrightarrow{E}\right| d_{+}\left| n_{0}\overrightarrow{E}%
\right\rangle -\left\langle n_{0}\overrightarrow{E}\right| d_{+}\left| l%
\overrightarrow{E}\right\rangle \left\langle l\overrightarrow{E}\right|
V_{w}\left| f\overrightarrow{E}\right\rangle \left\langle f\overrightarrow{E}%
\right| d_{-}\left| n_{0}\overrightarrow{E}\right\rangle \right] }{%
E_{f}-E_{l}}\right\} $

\noindent and
\begin{eqnarray}
\beta _{split}^{T} &=&\frac{\omega }{\hbar }\sum_{N_{0}}P(N_{0})\sum_{F}%
\left\{ \frac{\left\langle n_{0}\right| d_{-}\left| f\right\rangle
\left\langle f\right| d_{+}\left| n_{0}\right\rangle -\left\langle
n_{0}\right| d_{+}\left| f\right\rangle \left\langle f\right| d_{-}\left|
n_{0}\right\rangle }{\omega _{FN_{0}}^{2}-\omega ^{2}}\right\} =  \nonumber
\\
&=&\frac{\omega }{\hbar }\sum_{N_{0}}P(N_{0})\sum_{F}\left\{ \frac{%
\left\langle n_{0}\right| d_{-}\left| f\right\rangle \left\langle f\right|
d_{+}\left| n_{0}\right\rangle -\left\langle n_{0}\right| d_{+}\left|
f\right\rangle \left\langle f\right| d_{-}\left| n_{0}\right\rangle }{%
(\omega _{FN_{0}}-\omega )(\omega _{FN_{0}}+\omega )}\right\} =  \nonumber \\
&=&\frac{1}{2\hbar }\sum_{N_{0}}P(N_{0})\sum_{F}\left\{ \frac{\left\langle
n_{0}\right| d_{-}\left| f\right\rangle \left\langle f\right| d_{+}\left|
n_{0}\right\rangle -\left\langle n_{0}\right| d_{+}\left| f\right\rangle
\left\langle f\right| d_{-}\left| n_{0}\right\rangle }{(\omega
_{FN_{0}}-\omega )}\right\}  \label{16-1}
\end{eqnarray}
\[
\omega _{FN_{0}}=\frac{E_{F}(\overrightarrow{E})-E_{N_{0}}(\overrightarrow{E}%
)}{\hbar }, 
\]
It should be reminded that according to all the above (see also section 3)
energy levels $E_{F}$ and $E_{N_{0}}$ contain shifts caused by interaction
of electric dipole moment of the level with electric field $\overrightarrow{E%
}$ and magnetic moment of the level with T-odd induced magnetic field $%
\overrightarrow{H}_{ind}(\overrightarrow{E})$.

It should be noted that radial parts of the atom wave functions are real 
\cite{9}, therefore the matrix elements of operators $d_{\pm }$ are real
too. As a result, the P-odd T-even part of the interaction $V_{w}$\ does not
contribute to $\beta _{mix}^{T}$\ \ because the P-odd T-even matrix elements
of $V_{w}$\ are imaginary \cite{7}.\ At the same time, the T- and P-odd
matrix elements of $V_{w}$\ are\ real \cite{7}, therefore, the
polarizability $\beta _{mix}^{T}\neq 0$. It should be mentioned that in the
absence of electric field ($\overrightarrow{E}=0$) the polarizability $\beta
_{E}^{T}=0$\ \ and, therefore, the phenomenon of the photon polarization
plane rotation is absent.

Really, the electric field $\overrightarrow{E}$\ mixes the opposite parity
levels of the atom . The atom levels have the fixed parity at $%
\overrightarrow{E}=0$. The operators\ $d_{\pm }$\ and $V_{w}$\ change the
parity of the atom states. As a result, the parity of the final state $%
\left| N_{0}^{\prime }\right\rangle =$\ \ $d_{+}$\ $d_{-}$\ $V_{w}$\ $\left|
N_{0}\right\rangle $\ appears to be opposite to the parity of the initial
state $\left| N_{0}\right\rangle $. But the initial and final states \ in
the expression for $\beta _{E}^{T}$\ are the same. Therefore $\beta _{E}^{T}$%
\ can not differ from zero\ at $\overrightarrow{E}=0$.

It should be emphasized once again that polarizability $\beta _{E}^{T}$\ \
differs from zero even if EDM of electron is equal to zero. The interaction
of electron EDM with electric field gives only part of contribution to the
total polarizability of atom (molecule). The new effect we discuss is caused
by the Stark mixing of atom (molecule) levels and weak T- and P-odd
interaction of electrons with nucleus (and with each other).

Therefore, according to (\ref{sum}) the total angle of polarization plane
rotation includes two terms $\vartheta =\vartheta _{mix}+\vartheta _{split}$%
, where \ $\vartheta _{mix}\sim \beta _{mix}^{T}$\ is caused by the
considered above effect similar to Van Vleck that and $\vartheta
_{split}\sim \beta _{split}^{T}$\ is caused by the \ atom levels splitting
both in electric field $\overrightarrow{E}$\ and magnetic field $%
\overrightarrow{H}_{ind}(\overrightarrow{E})$. The contributions given by $%
\beta _{mix}^{T}$\ and \ $\beta _{split}^{T}$\ can be distinguished by
studying the frequency dependence of $\vartheta =\vartheta (\omega )$.
According to (\ref{16},\ref{16-1}) $\vartheta _{mix}\sim \frac{1}{\omega
_{fn_{0}}-\omega }$\ whereas $\vartheta _{split}\sim \frac{1}{(\omega
_{fn_{0}}-\omega )^{2}}.$\ So, $\vartheta _{split}$\ \ decreases faster then 
$\vartheta _{mix}$ with the grows of frequency tuning out from resonance.

Furthermore there is one more possibility to distinguish contribution of $%
\overrightarrow{d_{a}}$\ and $\overrightarrow{H}_{ind}(\overrightarrow{E})$\
from that given by $\beta _{E}^{T}.$\ This can be done when the photon beam
is incident orthogonally to the electric field $\overrightarrow{E}$. As it
was shown above, the T-odd contribution to birefringence effect (depending
on $\overrightarrow{d_{a}}$\ and $\overrightarrow{H}_{ind}(\overrightarrow{E}%
)$)\ appears in this case. Only the symmetrical part of {\em \ }tensor of
dynamical polarizability $\alpha _{ik}$\ of an atom (molecule){\em \ }(\ref
{11}) contributes in it and antisymmetric one does not. \ To observe the
T-odd birefringence effect it's more convenient to study atom exposed both
to electric and magnetic fields. In this case the effect magnitude is
proportional to $(\overrightarrow{H}\overrightarrow{E})$\ and one can easily
pick out T-odd effect among T-even ones changing$\overrightarrow{E}$\
direction with respect to $\overrightarrow{H}.$\ It should be noted that two
effects contribute in T-odd birefringence: 1. levels splitting, 2. mixing of
ground state and excited state energy levels of atom in external fields.

Let us now estimate the magnitude of the effect of the T-odd photon plane
rotation due to $\beta _{mix}^{T}$. According to the analysis \cite{4,5,6},
based on the calculations \ of the value of \ T- and P-noninvariant
interactions given by \cite{7}, the ratio $\frac{\left\langle
V_{w}^{T}\right\rangle }{\left\langle V_{w}^{P}\right\rangle }\leq
10^{-3}\div 10^{-4}$, where\ $\left\langle V_{w}^{T}\right\rangle $\ is T
and P-odd matrix element, $\left\langle V_{w}^{P}\right\rangle $\ is P-odd
T-even matrix element.

The P-odd T-even polarizability $\beta _{s}^{P}$\ is proportional to the
electric dipole matrix element, the magnetic dipole matrix element and $%
\left\langle V_{w}^{P}\right\rangle $: $\beta _{s}^{P}\sim $\ $\left\langle
d\right\rangle \left\langle \mu \right\rangle \left\langle
V_{w}^{P}\right\rangle $\ \cite{7}. At the same time $\beta _{mix}^{T}\sim $%
\ $\left\langle d(\overrightarrow{E})\right\rangle \left\langle d(%
\overrightarrow{E})\right\rangle \left\langle V_{w}^{T}\right\rangle $. As a
result, 
\begin{equation}
\frac{\beta _{mix}^{T}}{\beta _{s}^{P}}\sim \frac{\left\langle d(%
\overrightarrow{E})\right\rangle \left\langle d(\overrightarrow{E}%
)\right\rangle \left\langle V_{w}^{T}\right\rangle }{\left\langle
d\right\rangle \left\langle \mu \right\rangle \left\langle
V_{w}^{P}\right\rangle }.  \label{17}
\end{equation}
Let us study the T-odd phenomena of the photon polarization plane rotation
in an electric field $\overrightarrow{E}$\ for the transition $%
n_{0}\rightarrow f$ \ between the levels $n_{0}$ and $f$\ which have the
same parity at $\overrightarrow{E}=0$.{\bf \ }The matrix element\ $%
\left\langle n_{0},\overrightarrow{E}\right| d_{\pm }\left| f,%
\overrightarrow{E}\right\rangle $\ does not equal to zero only if $%
\overrightarrow{E}\neq 0$ . Let the energy of interaction of an atom with an
electric field, $V_{E}=-$ $\overrightarrow{d}\overrightarrow{E},$\ be much
smaller than the spacing $\Delta $\ of the energy levels, which are mixed by
the field $\overrightarrow{E}$. Then one can use the perturbation theory for
the wave functions $\left| f,\overrightarrow{E}\right\rangle $: 
\begin{equation}
\left| f,\overrightarrow{E}\right\rangle =\left| f\right\rangle +\sum_{m}%
\frac{\left\langle m\right| -d_{z}E\left| f\right\rangle }{E_{f}-E_{m}}%
\left| m\right\rangle ,  \label{18}
\end{equation}
where $z\parallel \overrightarrow{E}$. As a result, the matrix element $%
\left\langle n_{0},\overrightarrow{E}\right| d_{\pm }\left| f,%
\overrightarrow{E}\right\rangle $\ \ can be rewritten as: 
\begin{eqnarray}
\left\langle n_{0},\overrightarrow{E}\right| d_{\pm }\left| f,%
\overrightarrow{E}\right\rangle &=&  \label{19} \\
&=&-\left\{ \sum_{m}\frac{\left\langle n_{0}\right| d_{\pm }\left|
m\right\rangle \left\langle m\right| d_{z}\left| f\right\rangle }{E_{f}-E_{m}%
}+\right.  \nonumber \\
&&+\left. \sum_{p}\frac{\left\langle n_{0}\right| d_{z}\left| p\right\rangle
\left\langle p\right| d_{\pm }\left| f\right\rangle }{E_{n_{0}}-E_{p}}%
\right\} E.  \nonumber
\end{eqnarray}
One can see that the matrix element $\left\langle d(\overrightarrow{E}%
)\right\rangle \sim \ \frac{\left\langle d\right\rangle E}{\Delta }$\ $%
\left\langle d\right\rangle $\ in this case. The other matrix elements in (%
\ref{16}) can be evaluated at $\overrightarrow{E}=0$. This gives the
estimate as follows: 
\begin{equation}
\beta _{mix}^{T}\sim \left\langle d\right\rangle \left\langle d\right\rangle 
\frac{\left\langle dE\right\rangle }{\Delta }\left\langle
V_{w}^{T}\right\rangle \ \ .  \label{20}
\end{equation}
and, consequently, ratio (\ref{17}) can be written as 
\begin{equation}
\frac{\beta _{mix}^{T}}{\beta _{s}^{P}}\sim \frac{\left\langle
d\right\rangle \left\langle d\right\rangle \frac{\left\langle
dE\right\rangle }{\Delta }\left\langle V_{w}^{T}\right\rangle }{\left\langle
d\right\rangle \left\langle \mu \right\rangle \left\langle
V_{w}^{P}\right\rangle }.  \label{21}
\end{equation}
Taking into account that the matrix element $\left\langle \mu \right\rangle
\sim \alpha \left\langle d\right\rangle $\ \ \cite{8,9}, where $\alpha =%
\frac{1}{137}$ is the fine structure constant, equation (\ref{21}) can be
reduced to: 
\begin{equation}
\frac{\beta _{mix}^{T}}{\beta _{s}^{P}}\sim \alpha ^{-1}\frac{\left\langle
dE\right\rangle }{\Delta }\frac{\left\langle V_{w}^{T}\right\rangle }{%
\left\langle V_{w}^{P}\right\rangle }  \label{22}
\end{equation}
For the case $\frac{\left\langle dE\right\rangle }{\Delta }\sim 1$, ratio (%
\ref{22}) gives 
\begin{equation}
\frac{\beta _{mix}^{T}}{\beta _{s}^{P}}\sim \alpha ^{-1}\frac{\left\langle
V_{w}^{T}\right\rangle }{\left\langle V_{w}^{P}\right\rangle }\lesssim
10^{-1}\div 10^{-2}  \label{23}
\end{equation}
Such condition can be realized, for example, for exited states of atoms and
for two-atom molecules (TlF, BiS, HgF) which have a pair of nearly
degenerate opposite parity states. As one can see, the ratio $\frac{\beta
_{mix}^{T}}{\beta _{s}^{P}}$\ is two orders larger as compared with the
simple estimation $\frac{\left\langle V_{w}^{T}\right\rangle }{\left\langle
V_{w}^{P}\right\rangle }\leq 10^{-3}\div 10^{-4}$\ due to the fact that $%
\beta _{mix}^{T}$\ is determined by only the electric dipole transitions ,
while $\beta _{s}^{P}$\ is determined by both the electric and magnetic
dipole transitions.

\newpage

{\bf {\Large 2. Time-violating photon polarization plane rotation by a
diffraction grating.}}

\bigskip

As it has been shown in \cite{4,5}, the energy of atom interaction with two
coherent electromagnetic waves depends on the T-violating scalar
polarizability $\beta _{t}^{T}$. Interaction of an atom (molecule) with two
waves can be considered as a process of re-scattering of one wave into
another and vice versa. Then, as it follows from an expression for the
effective interaction energy, the amplitude $f\left( \vec{k}^{^{\prime }},%
\vec{k}\right) $ of the photon scattering by an unpolarized atom (molecule)
at a non-zero angle is given by \cite{5}:
\begin{equation}
f\left( \vec{k}^{^{\prime }},\vec{k}\right) =f_{ik}e_{i}^{^{\prime ^{\ast
}}}e_{k}=\frac{\omega ^{2}}{c^{2}}\left( \alpha _{s}\vec{e}^{^{\prime ^{\ast
}}}\vec{e}+i\frac{1}{2}\beta _{s}^{P}\left( \vec{n}^{^{\prime }}+\vec{n}%
\right) \left[ \vec{e}^{^{\prime ^{\ast }}}\vec{e}\right] +\beta
_{s}^{T}\left( \vec{n}^{^{\prime }}-\vec{n}\right) \left[ \vec{e}^{^{\prime
^{\ast }}}\vec{e}\right] \right) ,  \label{2-7}
\end{equation}
where $\vec{k}$ is the wave vector of a scattered photon, $\vec{n}=\frac{\vec{k}}{k}$, 
$\vec{n}^{^{\prime
}}=\frac{\vec{k}^{^{\prime }}}{k}$ , $\alpha _{s}$ is the scalar
P,T-invariant polarizability of an atom (molecule). Expression (\ref{2-7})
holds true in the absence of external electric and magnetic fields.

The elastic scattering amplitude (\ref{2-7}) can be derived from the general
principles of symmetry. Indeed, there are four independent unit vectors: $%
\vec{\nu}_{1}=\frac{\vec{k}^{^{\prime }}+\vec{k}}{\left| \vec{k}^{^{\prime
}}+\vec{k}\right| }$ , $\vec{\nu}_{2}=\frac{\vec{k}^{^{\prime }}-\vec{k}}{%
\left| \vec{k}^{^{\prime }}-\vec{k}\right| }$ , $\vec{e}$ and $\vec{e}^{^{\
\prime }}$ , which completely describe geometry of the elastic scattering
process. The elastic scattering amplitude $f\left( \vec{k}^{^{\prime }},\vec{%
k}\right) $ depends on these vectors and therewith is a scalar. Obviously,
one can compose three independent scalars from these vectors: $(\vec{e}^{^{\
\prime }}\vec{e})$ , $(\vec{\nu}_{1}\left[ \vec{e}^{^{\ \prime ^{\ast }}}%
\vec{e}\right] )$, $(\vec{\nu}_{2}\left[ \vec{e}^{^{\ \prime ^{\ast }}}\vec{e%
}\right] )$. As a result, the scattering amplitude can be written as: 
\begin{equation}
f\left( \vec{k}^{^{\prime }},\vec{k}\right) =f_{s}\left( \vec{k}^{^{\prime
}},\vec{k}\right) \vec{e}^{^{\ \prime ^{\ast }}}\vec{e}+if_{s}^{P}\left( 
\vec{k}^{^{\prime }},\vec{k}\right) \vec{\nu}_{1}\left[ \vec{e}^{^{\ \prime
^{\ast }}}\vec{e}\right] +f_{s}^{T}\left( \vec{k}^{^{\prime }},\vec{k}%
\right) \vec{\nu}_{2}\left[ \vec{e}^{^{\ \prime ^{\ast }}}\vec{e}\right] ,
\label{2-8}
\end{equation}
where $f_{s}$ is the P-,T- invariant scalar amplitude, $f_{s}^{P}$ is the
P-violating scalar amplitude, and $f_{s}^{T}$ is the P-,T- violating scalar
amplitude.

It can easily be found from (\ref{2-7},\ref{2-8}) that the term proportional
to $\beta _{s}^{T}\left( f_{s}^{T}\right) $ vanishes in the case of forward
scattering $\left( \vec{n}^{^{\prime }}\rightarrow \vec{n}\right) $. Vice
versa, in the case of back scattering $\left( \vec{n}^{^{\prime
}}\rightarrow -\vec{n}\right) $ the term proportional to $\beta
_{s}^{P}\left( f_{s}^{P}\right) $ gets equal to zero.

Thus, one can conclude that the T-violating interactions manifest themselves
in the processes of scattering by atoms (molecules). However, the scattering
processes are usually incoherent and their cross sections are too small to
hope for observation of the T-violating effect. Another situation takes
place for diffraction gratings in the vicinity of the Bragg resonance where
the scattering process is coherent. As a result, the intensities of
scattered waves strongly increase: for instance, in the Bragg (reflection)
diffraction geometry the amplitude of the diffracted-reflected wave may
reach the unity. It gives us an opportunity to study the T-violating
scattering processes \cite{5} (the detailed discussion see at \cite{winter-school}).

To include the P, T violating processes into the diffraction theory, let us
consider the microscopic Maxwell equations:
\begin{eqnarray}
curl\vec{E} &=&-\frac{1}{c}\frac{\partial \vec{B}}{\partial t}\quad ,\quad
curl\vec{B}=\frac{1}{c}\frac{\partial \vec{E}}{\partial t}+\frac{4\pi }{c}%
\vec{j}\quad ,  \label{2-9} \\
div\vec{E} &=&4\pi \rho \quad ,\quad div\vec{B}=0\quad ,\quad \frac{\partial
\rho }{\partial t}+div\vec{j}=0\quad .  \nonumber
\end{eqnarray}
where $\vec{E}$ is the electric field strength and $\vec{B}$ is the magnetic
field induction, $\rho $ and $\vec{j}$ are the microscopic densities of the
electrical charge and the current induced by an electromagnetic wave, $c$ is
the speed of light. The Fourier transformation of these equations (i.e. $%
\vec{E}\left( \vec{r},t\right) =\frac{1}{2\pi ^{4}}\int \vec{E}\left( \vec{k}%
,\omega \right) e^{i\vec{k}\vec{r}}e^{-i\omega t}d^{3}kd\omega $ and so on)
yields to equation for $\vec{E}\left( \vec{k},\omega \right) $:
\begin{equation}
\left( -k^{2}+\frac{\omega ^{2}}{c^{2}}\right) {\small \vec{E}}\left( \vec{k}%
,\omega \right) {\small =-}\frac{4\pi i\omega }{c^{2}}\left[ \vec{j}\left( 
\vec{k},\omega \right) -\frac{c^{2}k^{2}}{\omega ^{2}}\vec{n}\left( \vec{n}%
\vec{j}\left( \vec{k},\omega \right) \right) \right] {\small ,}\;
\label{2-10}
\end{equation}
where $\vec{n}=\frac{\vec{k}}{k}$ .

In linear approximation, the current $\vec{j}\left( \vec{r},\omega \right) $
is coupled with $\vec{E}\left( \vec{r},\omega \right) $ by the well-known
dependence: $j_{i}\left( \vec{r},\omega \right) =\int d^{3}r^{^{\prime
}}\sigma _{ij}\left( \vec{r},\vec{r}^{^{\prime }},\omega \right) E_{j}\left( 
\vec{r}^{^{\prime }},\omega \right) $ with $\sigma _{ij}\left( \vec{r},\vec{r%
}^{^{\prime }},\omega \right) $ as the microscopic conductivity tensor being
a sum of the conductivity tensors of the atoms (molecules) constituting the
diffraction grating: $\sigma _{ij}\left( \vec{r},\vec{r}^{^{\prime }},\omega
\right) =\sum_{A=1}^{N}\sigma _{ij}^{A}\left( \vec{r},\vec{r}^{^{\prime
}},\omega \right) ,$ here $\sigma _{ij}^{A}$ is the conductivity tensor of
the A-type scatterers. The summation is done over all atoms (molecules) of
the grating. In a diffraction grating, the tensor $\sigma _{ij}\left( \vec{r}%
,\vec{r}^{^{\prime }},\omega \right) $ is a spatially periodic function.
Therefore, $j_{i}\left( \vec{k},\omega \right) $ can be written as follows:
\begin{equation}
j_{i}\left( \vec{k},\omega \right) =\frac{1}{V_{0}}\sum_{\vec{\tau}}\sigma
_{ij}^{c}\left( \vec{k},\vec{k}-\vec{\tau},\omega \right) E_{j}\left( \vec{k}%
-\vec{\tau},\omega \right)  \label{2-11}
\end{equation}
where $\sigma _{ij}^{c}$ is the Fourier transform of the conductivity tensor
of a grating's elementary cell, $\vec{\tau}$ is the reciprocal lattice
vector of the diffraction grating.

Using current representation (\ref{2-11}), one can obtain from (\ref{2-10}): 
\begin{equation}
\left( -k^{2}+k_{0}^{2}\right) E_{i}\left( \vec{k},\omega \right) =-\frac{%
\omega ^{2}}{c^{2}}\sum_{\vec{\tau}}\hat{\chi}_{ij}\left( \vec{k},\vec{k}-%
\vec{\tau}\right) E_{j}\left( \vec{k}-\vec{\tau}\right)  \label{2-12}
\end{equation}
Tensor of the diffraction grating susceptibility is given by 
\begin{equation}
\hat{\chi}_{ij}\left( \vec{k},\vec{k}-\vec{\tau}\right) =\left( \delta
_{il}-n_{i}n_{l}\right) \chi _{lj}\left( \vec{k},\vec{k}-\vec{\tau}\right)
\label{2-13}
\end{equation}
with 
\[
\chi _{lj}\left( \vec{k},\vec{k}-\vec{\tau}\right) =\frac{4\pi i}{%
V_{0}\omega }\sigma _{lj}\left( \vec{k},\vec{k}-\vec{\tau}\right) =\frac{%
4\pi c^{2}}{V_{0}\omega ^{2}}F_{lj}\left( \vec{k},\vec{k}-\vec{\tau}\right)
. 
\]
Here $F_{lj}\left( \vec{k},\vec{k}-\vec{\tau}\right) =\frac{i\omega }{c^{2}}%
\sigma _{lj}\left( \vec{k},\vec{k}-\vec{\tau}\right) $ is the amplitude of
coherent elastic scattering of an electromagnetic wave by a grating
elementary cell from a state with the wave vector $(\vec{k}-\vec{\tau})$ to
a state with the wave vector $\vec{k}$.

The amplitude $F_{lj}$ is obtained by summation of atomic (molecular)
coherent elastic scattering amplitudes over a grating's elementary cell:
\begin{equation}
F_{lj}\left( \vec{k}^{^{\prime }}=\vec{k}+\vec{\tau},\vec{k}\right)
=\left\langle \sum_{A=1}^{N_{c}}f_{lj}^{A}\left( \vec{k}^{^{\prime }}=\vec{k}%
+\vec{\tau},\vec{k}\right) e^{-i\vec{\tau}\vec{R}_{A}}\right\rangle ,
\label{2-14}
\end{equation}
where $f_{lj}^{A}$ is the coherent elastic scattering amplitude by an A-type
atom (molecule), $\vec{R}_{A}$ is the gravity center coordinate of the
A-type atom (molecule) , $N_{c}$ is the number of the atoms (molecules) in
an elementary cell, angular brackets denote averaging over the coordinate
distribution of scatterers in a grating's elementary cell. The amplitude $%
f_{lj}$ is given by equation (\ref{2-7},\ref{2-8}).

From (\ref{2-13}), (\ref{2-14}) and (\ref{2-8}) one can obtain an expression
for the susceptibility $\chi _{lj}$ of the elementary cell of an optically
isotropic material:
\begin{equation}
\chi _{lj}\left( \vec{k},\vec{k}-\vec{\tau}\right) =\chi _{s\vec{\tau}%
}\delta _{lj}+i\chi _{s\vec{\tau}}^{P}\varepsilon _{ljf}\nu _{1f}^{\vec{\tau}%
}+\chi _{s\vec{\tau}}^{T}\varepsilon _{ljf}\nu _{2f}^{\vec{\tau}}
\label{2-15}
\end{equation}
where 
\[
\chi _{s\vec{\tau}}^{(P,T)}=\frac{4\pi c^{2}}{V_{0}\omega ^{2}}\left\langle
\sum_{A=1}^{N_{c}}f_{s}^{A(P,T)}\left( \vec{k},\vec{k}-\vec{\tau}\right)
e^{-i\vec{\tau}\vec{R}_{A}}\right\rangle , 
\]
$\chi _{s\vec{\tau}}$ is the scalar P-, T- invariant susceptibility of an
elementary cell, $\chi _{s\vec{\tau}}^{P}$ is the P-violating, T- invariant
susceptibility of the elementary cell, and $\chi _{s\vec{\tau}}^{T}$ is the
P- and T- violating susceptibility of the elementary cell, 
\[
\vec{\nu}_{1}^{\vec{\tau}}=\frac{2\vec{k}-\vec{\tau}}{\left| 2\vec{k}-\vec{%
\tau}\right| }\quad ,\quad \vec{\nu}_{2}^{\vec{\tau}}=\frac{\vec{\tau}}{\tau 
} 
\]

Then, using (\ref{2-12},\ref{2-13},\ref{2-15}) one can derive a set of
equations describing the P and T violating interaction of an electromagnetic
wave with a diffraction grating
\begin{eqnarray}
\left( -\frac{k^{2}}{k_{0}^{2}}+1\right) E_{i}\left( \vec{k}\right)
&=&-\left( \delta _{ij}-n_{i}n_{j}\right) \chi _{_{s0}}E_{j}\left( \vec{k}%
\right) -i\chi _{s0}^{P}\left( \delta _{il}-n_{i}n_{l}\right) \varepsilon
_{ljf}n_{f}E_{j}\left( \vec{k}\right) -  \nonumber \\
&&-\sum_{\vec{\tau}\neq 0}\{\left( \delta _{ij}-n_{i}n_{j}\right) \chi _{_{s%
\vec{\tau}}}E_{j}\left( \vec{k}-\vec{\tau}\right) +  \nonumber \\
&&+i\chi _{_{s\vec{\tau}}}^{P}\left( \delta _{il}-n_{i}n_{l}\right)
\varepsilon _{ljf}\nu _{1f}^{\vec{\tau}}E_{j}\left( \vec{k}-\vec{\tau}%
\right) +  \label{2-16} \\
&&+\chi _{s_{\vec{\tau}}}^{T}\left( \delta _{il}-n_{i}n_{l}\right)
\varepsilon _{ljf}\nu _{2f}^{\vec{\tau}}E_{j}\left( \vec{k}-\vec{\tau}%
\right) \},  \nonumber
\end{eqnarray}
where $k_{0}=\frac{\omega }{c}$

Assuming the interaction to be P, T invariant $\left( \chi _{s}^{P}=\chi
_{s}^{T}=0\right) $, eqns. (\ref{2-16}) are reduced to the conventional set
of equations of dynamic diffraction theory\ \cite{chang}. \bigskip The
detailed analysis of these equations was done in \cite{6}. \ \qquad

According to \cite{6} the angle of the photon polarization plane rotation
out of Bragg conditions is defined by
\begin{equation}
\vartheta =-k_{0}Re\chi _{s}^{P}\left( 0\right) L+2k_{0}\alpha _{\tau
}^{-1}Re\left[ \chi _{1s}\left( \vec{\tau}\right) \chi _{2s}^{T}\left( \vec{%
\tau}\right) -\chi _{2s}\left( \vec{\tau}\right) \chi _{1s}^{T}\left( \vec{%
\tau}\right) \right] L  \label{2-17}
\end{equation}
So, the T-violating rotation arises in the case of nonzero odd part of the
susceptibility: $\chi _{2}\left( \vec{\tau}\right) \neq 0$. Such a situation
is possible if an elementary cell of the diffraction grating does not posses
the center of symmetry.

In accordance with (\ref{2-17}), the angle of the T-violating rotation grows
at $\alpha _{\tau }\rightarrow 0$. However, the condition $\alpha _{\tau
}\left| \chi _{s}\left( \vec{\tau}\right) \right| \ll 1$ is violated at $%
\alpha _{\tau }^{-1}\rightarrow 0$, when the amplitude of diffracted and
transmitted waves are comparable: $E\left( \vec{k}-\vec{\tau}\right) \simeq
E\left( \vec{k}\right) $ and, consequently, the perturbation theory gets
unapplicable. A rigorous dynamical diffraction theory should be applied in
this case.

Let the Bragg condition is fulfilled only for the diffracted wave. It allows
us to use the two-wave approximation of the dynamical diffraction theory 
\cite{chang}. Then, the set of equations (\ref{2-10}) is reduced to two
coupled equations, which for the back-scattering diffraction scheme $\left( 
\vec{k}_{0}\parallel \vec{\tau}\right) $ take the form \cite{6}
\begin{eqnarray}
\left( \frac{k^{2}}{k_{0}^{2}}-1\right) E_{j}\left( \vec{k}\right) &=&\chi
_{s}\left( 0\right) E_{j}\left( \vec{k}\right) +i\chi _{s}^{P}\left(
0\right) \varepsilon _{jmf}E_{m}\left( \vec{k}\right) n_{f}+  \nonumber \\
&&+\chi _{s}\left( \vec{\tau}\right) E_{j}\left( \vec{k}-\vec{\tau}\right)
+\chi _{s}^{T}\left( \vec{\tau}\right) \varepsilon _{jmf}E_{m}\left( \vec{k}-%
\vec{\tau}\right) \nu _{2f}^{\vec{\tau}},  \nonumber \\
&&  \label{2-18} \\
\left( \frac{\left( \vec{k}-\vec{\tau}\right) ^{2}}{k_{0}^{2}}-1\right)
E_{j}\left( \vec{k}-\vec{\tau}\right) &=&\chi _{s}\left( 0\right)
E_{j}\left( \vec{k}-\vec{\tau}\right) +  \nonumber
\end{eqnarray}

\[
i\chi _{s}^{P}\left( 0\right) \varepsilon _{jmf}n_{f}\left( \vec{k}-\vec{\tau%
}\right) E_{m}\left( \vec{k}-\vec{\tau}\right) +\chi _{s}\left( -\vec{\tau}%
\right) E_{j}\left( \vec{k}\right) +\chi _{s}^{T}\left( -\vec{\tau}\right)
\varepsilon _{jmf}\nu _{2f}^{-\vec{\tau}}E_{m}\left( \vec{k}\right) , 
\]
$\vec{n}\left( \vec{k}-\vec{\tau}\right) =\frac{\vec{k}-\vec{\tau}}{\left| 
\vec{k}-\vec{\tau}\right| }$ .

These set of equations can be diagonalized for the photon with a certain
circular polarization. Let the right-circularly polarized photon $\left( 
\vec{e}_{+}\right) $ be incident on the diffraction grating. Then, the
diffraction process yields to the appearance of a back-scattered photon with
the left circular polarization $\left( \vec{e}_{-}^{\vec{\tau}}\right) $
(this is because the momentum of the back-scattered photon $\vec{k}%
^{^{\prime }}=\vec{k}-\vec{\tau}$ is antiparallel to the momentum $\vec{k}$
of the incident one). And visa versa, the left-circularly polarized photon
produces a right-circularly polarized back-scattered one.

Thus, for circularly polarized photons the set of vector equations (\ref
{2-18}) can be split into two independent sets of scalar equations \cite{6}.
The explicit solution of these equations yield to the following expression
for the transmitted wave amplitude \cite{6} (all the symbols are defined in 
\cite{6}):
\[
\vec{E}_{\pm }=\vec{e}_{\pm }\left( -1\right) ^{b}e^{i\varphi _{\pm }},\;%
\text{where }\varphi _{\pm }=k_{0}\left[ \frac{1}{2}\varepsilon ^{\pm
}\left( \alpha _{1,2}\right) -\frac{k_{0}\left( \alpha _{1,2}-2\chi
_{s}\left( 0\right) \right) L}{8\pi b}\Delta ^{\pm }\right] L 
\]
Using this equation one can find the angle of the polarization plane rotation
\begin{center}
$\vartheta =Re\left( \varphi _{+}-\varphi _{-}\right) =\vartheta
^{P}+\vartheta _{1,2}^{T},\;$\ 
\end{center}
where $\vartheta ^{P}=-k_{0}Re\chi _{s}^{P}\left( 0\right) L$ - defines the
P-violating T-invariant rotation angle and $\vartheta _{1,2}^{T}$
corresponds to the T-violating rotation:
\begin{eqnarray}
\vartheta _{1,2}^{T}\left( \alpha _{1,2}\right) &=&\mp \frac{k_{0}^{3}L^{3}}{%
8\pi ^{2}b^{2}}\sqrt{4\left( \chi _{1s}^{2}+\chi _{2s}^{2}\right) +\left( 
\frac{4\pi b}{k_{0}L}\right) ^{2}}\times  \label{2-19} \\
&&\times \left[ \chi _{1s}\left( \vec{\tau}\right) Re\chi _{2s}^{T}\left( 
\vec{\tau}\right) -\chi _{2s}\left( \vec{\tau}\right) Re\chi _{1s}^{T}\left( 
\vec{\tau}\right) \right] ,  \nonumber
\end{eqnarray}
the sign $\left( -\right) $ corresponds to $\alpha _{1}$ , the sign $\left(
+\right) $ corresponds to $\alpha _{2}$.

The imaginary part of the T-violating polarizability $Im\chi _{s1,2}^{T}$ is
responsible for the T-violating circular dichroism. Due to that process, a
linearly polarized photon gets a circular polarization at the diffraction
grating's output. The degree of the circular polarization of the photon is
determined by the relation:
\begin{equation}
\delta _{1,2}=\frac{\left| \vec{E}_{+}\right| ^{2}-\left| \vec{E}_{-}\right|
^{2}}{\left| \vec{E}_{+}\right| ^{2}+\left| \vec{E}_{-}\right| ^{2}}\simeq
Im\varphi _{-}-Im\varphi _{+}=k_{0}Im\chi _{s}^{P}\left( 0\right) L\pm
\label{2-20}
\end{equation}

\[
\pm \frac{k_{0}^{3}L^{3}}{8\pi ^{2}b^{2}}\sqrt{4\left( \chi _{1s}^{2}+\chi
_{2s}^{2}\right) +\left( \frac{4\pi b}{k_{0}L}\right) ^{2}}\left[ \chi
_{1s}\left( \vec{\tau}\right) Im\chi _{2s}^{T}\left( \vec{\tau}\right) -\chi
_{2s}\left( \vec{\tau}\right) Im\chi _{1s}^{T}\left( \vec{\tau}\right) %
\right] 
\]
It should be pointed out that the resonance transmission condition is
satisfied at a given $b$ for two different values of $\alpha $. This is
because there is a possibility to approach to the Brilluan (the total Bragg
reflection) bandgap both from high and low frequencies. The T-violating
parts of the rotation angle are opposite in sign for $\alpha _{1}$ and for $%
\alpha _{2}$. It gives the addition opportunity to distinguish the
T-violating rotation from the P-violating T-invariant rotation. Indeed, the
P-violating rotation does not depend on the back Bragg diffraction in the
general case because the P-violating scattering amplitude equals to zero for
back scattering (see \cite{6}). In accordance with (19,20) the T-violating
rotation and dichroism grow sharply in the vicinity of the resonance Bragg
transmission. At the first view, one could expect for $\vartheta ^{T}$ the
dependence $\vartheta ^{T}\sim k_{0}Re\chi _{s1,2}^{T}\left( \vec{\tau}%
\right) L$ (see (\ref{2-17})). However, in the vicinity of resonance, the
rotation angle $\vartheta ^{T}$ turns out to be multiplied by the factor $%
B=(8\pi ^{2}b^{2})^{-1}k_{0}\sqrt{4\left( \chi _{1s}^{2}+\chi
_{_{2s}}^{2}\right) +\left( \frac{4\pi b}{k_{0}L}\right) ^{2}}Lk_{0}\chi
_{s1,2}L$ which provides the above mentioned growth (for example, $B\sim
10^{5}$ at $\chi _{s}\simeq 10^{-1}$ , $k_{0}\simeq 10^{4}\div 10^{5}cm^{-1}$
, $L=1cm$ , $b=1$ ).

Now, let us estimate the effect magnitude. To do this we must determine (in
accordance with (\ref{2-19})) the T-violating susceptibility $\chi
_{s1,2}^{T}$, which is proportional to the T-violating polarizability $\beta
_{s}^{T}$ of atom. The estimate carried out by \cite{4,5,7} gives $\beta
_{s}^{T}\sim 10^{-3}\div 10^{-4}\beta _{s}^{P}$, where $\beta _{s}^{P}$ is
the P-violating T-invariant scalar polarizability. The polarizability $\beta
_{s}^{P}$ was studied both theoretically and experimentally \cite{7}.
Particularly, the theory gives $\beta _{s}^{P}\cong 10^{-30}cm^{3}$ for
atoms analogous to Bi, Tl, Pb. It yields to the estimate $\cong 10^{-33}\div
10^{-34}cm^{3}$ for the T-violating atomic polarizability. The
polarizability $\beta _{s}^{P}$ causes the P-violating rotation of the
polarization plane by the angle $\vartheta ^{P}=kRe\chi _{s}^{P}\left(
0\right) L\cong 10^{-7}$ rad/cm$\times $L for the gas density $\rho
=10^{16}\div 10^{17}cm^{-3}$. As a result, in our case the parameter $%
\varphi =k\chi _{s}^{T}\left( \tau \right) L$ turns out to be $\varphi \cong
10^{-10}\div 10^{-11}$ rad/cm$\times $L and can be even less by the factor $%
h/d$, where $h$ is the corrugation amplitude of the diffraction grating
while $d$ is the distance between waveguide's mirrors. Assuming this factor
to be ~$\sim 10^{-1}$, we shall find $\varphi \cong 10^{-11}\div 10^{-12}$
rad/cm$\times $L. Thus, the final estimate of the T-violating rotation angle 
$\vartheta ^{T}$ is
\begin{equation}
\vartheta ^{T}\cong 10^{-11}\div 10^{-12}\frac{rad}{cm}{k_{0}}^{2}{\chi
_{s}^{2}\left( \tau \right) }L^{3}  \label{2-21}
\end{equation}
In real situation the susceptibility of a grating $\chi _{s}\left( \tau
\right) $ can exceed the unity. However, our analysis has been performed
under the assumption $\chi _{s}\ll 1$. Suppose $\chi _{s}=10^{-1}$ , $%
k_{0}=10^{4}$ then $\vartheta ^{T}\simeq 10^{-6}\div 10^{-7}L^{3}$ and,
consequently, for L= 1 cm one can get the rotation angle $\vartheta
^{T}\simeq 10^{-6}\div 10^{-7}rad$.

Obviously, the obtained the T-violating rotation angle $\vartheta ^{T}$ is
of the same order as compared with $\vartheta ^{P}$. It makes possible
experimental observation of the phenomenon of the T-violating polarization
plane rotation.

It should be noted that the manufacturing of diffraction gratings for the
wave lengths longer than visible light can be simpler. That is why we would
like to attract attention to the possibility of studying of the T-violating
polarization plane rotation in the vicinity of frequencies of atomic
(molecular) hyperfine transitions; for example, for Ce (the transition
wavelength is $\lambda =3.26$ cm) and Tl ( $\lambda =1.42$ cm).

Thus, we have shown that the phenomenon of the T-violating polarization
plane rotation appears when the photon is scattered by a volume diffraction
grating. The phenomenon grows sharply in the vicinity of the resonance
transmission condition. An experimental scheme based on a waveguide,
containing a diffraction grating and gas, has been proposed that enables
real experiments on observation of the T-violating polarization plane
rotation to be performed. The rotation angle has been shown to be $\vartheta
^{T}=10^{-6}\div 10^{-7}L^{3}$, where L is the waveguide length (thickness
of the equivalent volume diffracting grating).

\newpage

{\bf {\Large 3. The possibility to observe the phenomena experimentally.}}

\bigskip

The possibility to observe the phenomena experimentally can be discussed
now. In accordance with (6) the angle of the T-odd rotation in electric
field can be evaluated as follows:
\begin{equation}
\vartheta _{mix}^{T}\sim \frac{2\pi \rho \omega }{c}\beta _{mix}^{T}L\sim 
\frac{\beta _{mix}^{T}}{\beta _{S}^{P}}\vartheta ^{P}\sim \alpha ^{-1}\frac{%
\left\langle dE\right\rangle }{\Delta }\frac{\left\langle
V_{w}^{T}\right\rangle }{\left\langle V_{w}^{P}\right\rangle }\vartheta ^{P}.
\label{2-22}
\end{equation}
According to the experimental data \cite{10,11} being well consistent with
calculations \cite{7} the typical value of $\vartheta ^{P}$\ is $\vartheta
^{P}\sim 10^{-6}rad$\ (for the length $L$\ being equal to the several
absorption lengths of the light propagating through a gas $L_{a}$).

For the electric field $E\sim 10^{4}V\cdot cm^{-1}$\ the parameter $\frac{%
\left\langle dE\right\rangle }{\Delta }$\ can be estimated as $\frac{%
\left\langle dE\right\rangle }{\Delta }\sim 10^{-5}$\ for Cs, Tl and $\frac{%
\left\langle dE\right\rangle }{\Delta }\sim 10^{-4}$\ for Yb and lead.
Therefore, one can obtain $\vartheta _{mix}^{T}\sim 10^{-13}rad$\ for Cs, Tl
and $\vartheta _{mix}^{T}\sim 10^{-12}rad$\ for Yb and lead. For the
two-atom molecules (TlF, BiS, HgF) the angle $\vartheta _{mix}^{T}$\ can be
larger, because they have a pair of degenerate opposite parity states.

It should be noted that the classical up-to-date experimental techniques
allow to measure angles of light polarization plane rotation up to $4,3\cdot
10^{-11}rad$\ \ \cite{tarasov}.

\begin{figure}[htbp]
\epsfxsize = 10 cm \centerline{\epsfbox{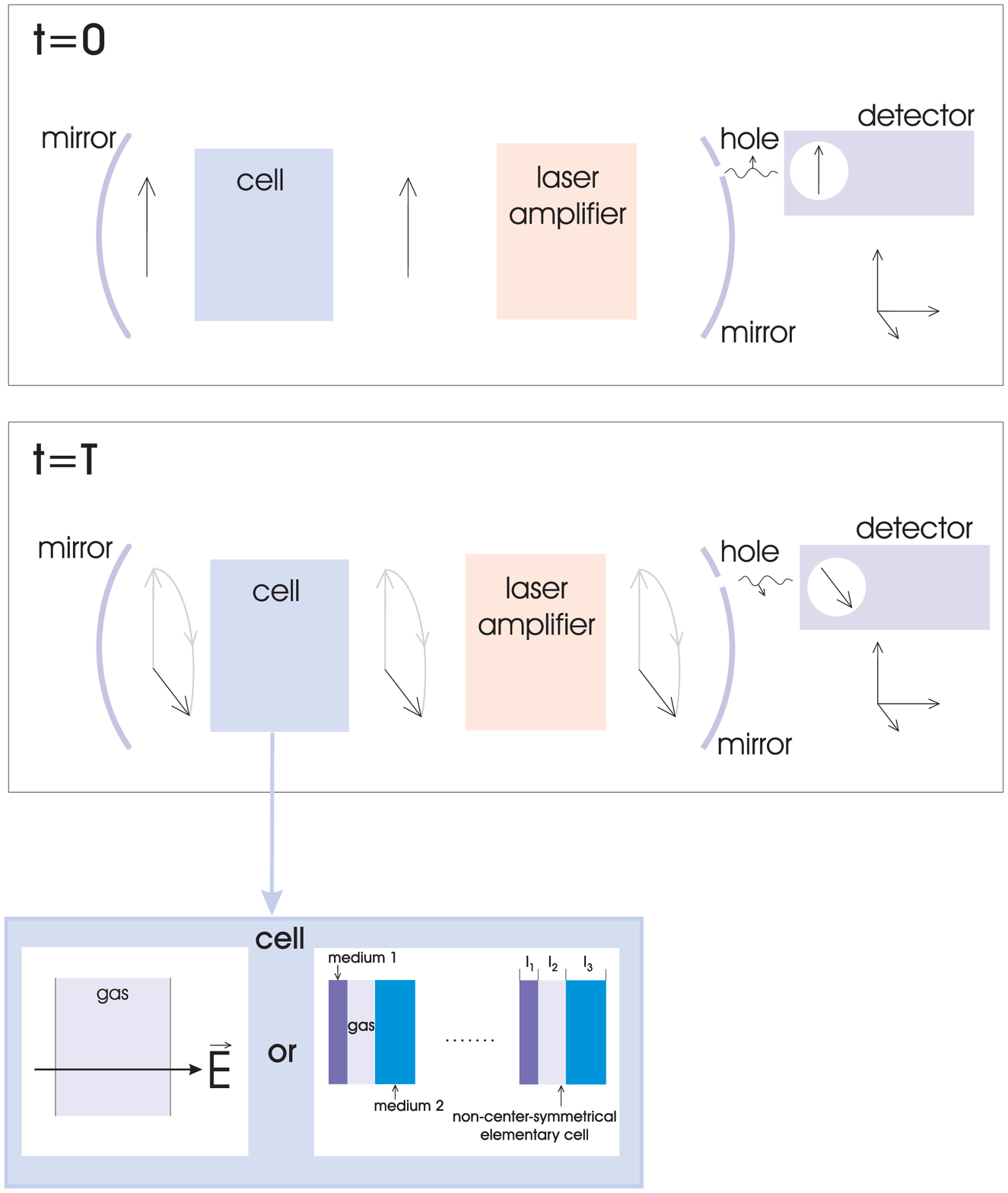}}
\caption{ }
\label{F1}
\end{figure}

\begin{figure}[htbp]
\epsfxsize = 7 cm \centerline{\epsfbox{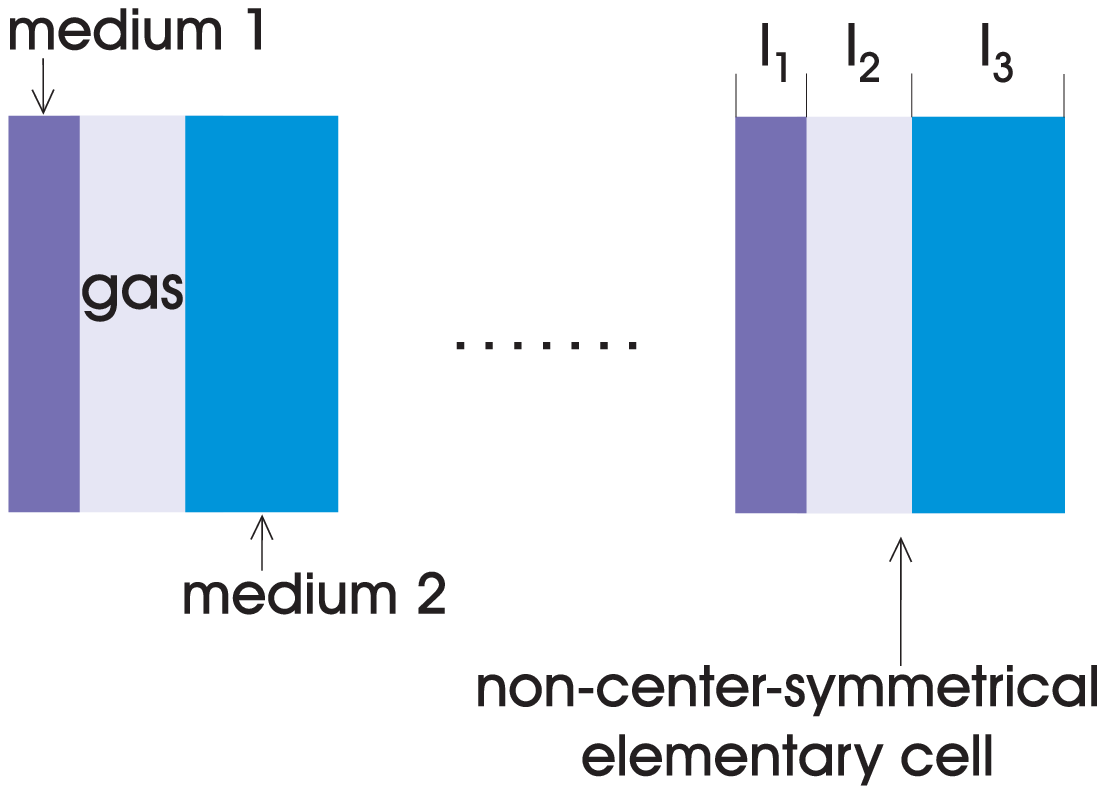}}
\caption{ }
\label{F2}
\end{figure}

\begin{figure}[htbp]
\epsfxsize = 10 cm \centerline{\epsfbox{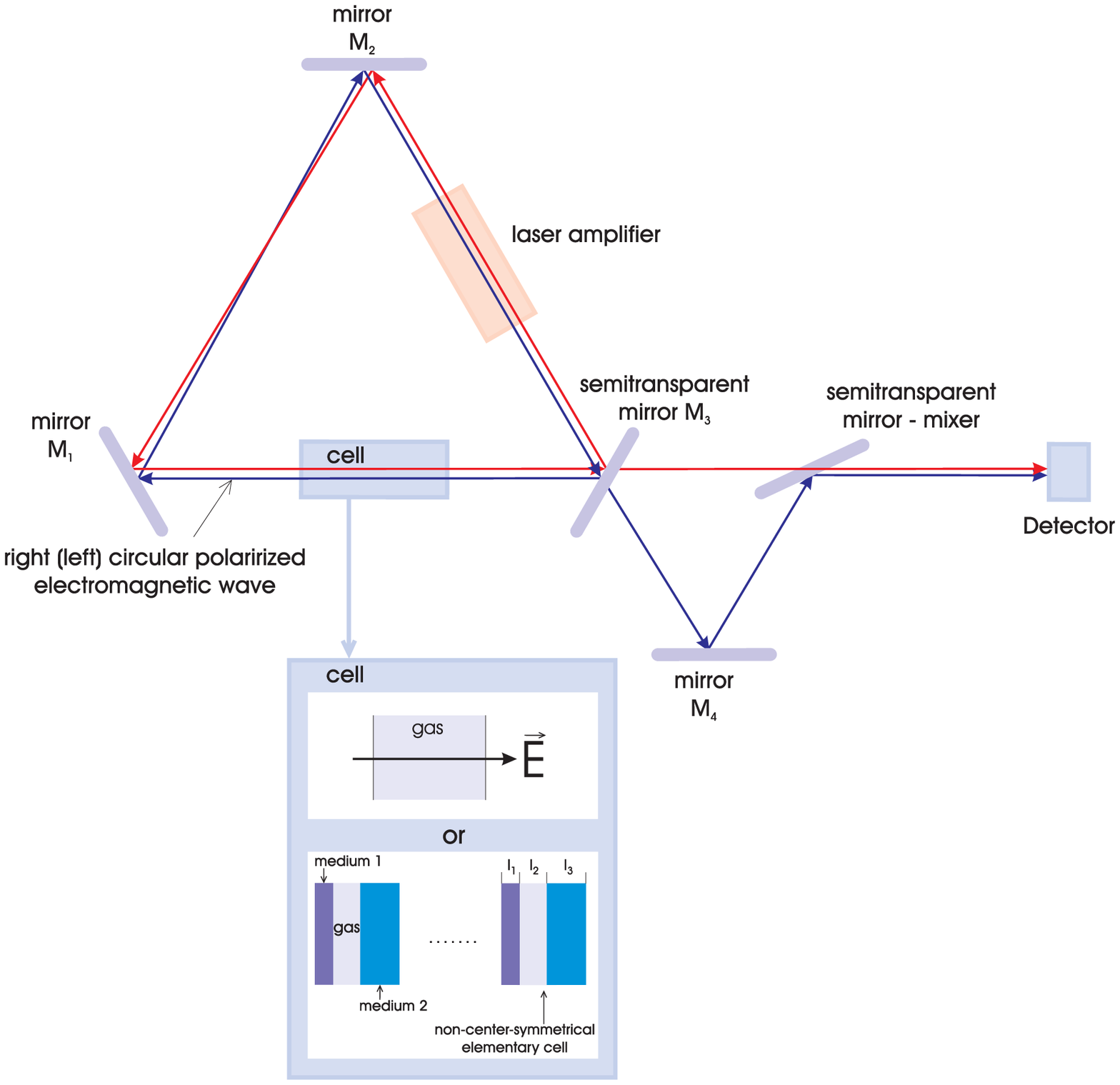}}
\caption{ }
\label{F3}
\end{figure}

\begin{figure}[htbp]
\epsfxsize = 10 cm \centerline{\epsfbox{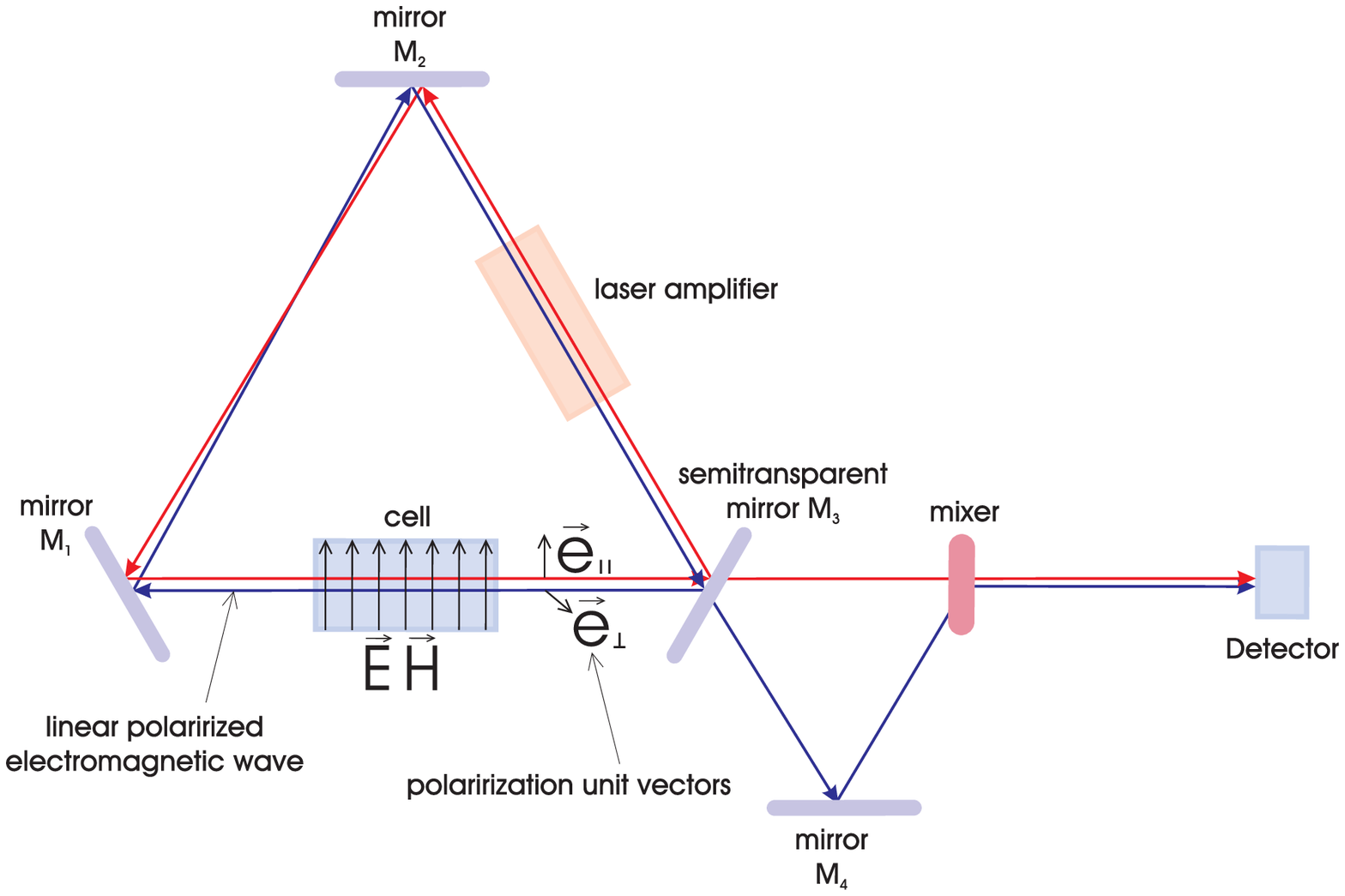}}
\caption{ }
\label{F4}
\end{figure}

A way to increase the rotation angle $\vartheta ^{T}$\ is to increase the
length $L$\ \ of the path of a photon inside a medium (see (6)). It can be
done, for example, by placing a medium (gas in an electric field or
non-center-symmetrical diffraction grating) in a resonator or inside a laser
gyroscope (Fig.\ref{F1}). This becomes possible due to the fact that in contrast
with the phenomenon of P-odd rotation of the polarization plane of photon
the T-odd rotation in an electric field (as well as in a diffraction
grating) is accumulated while photon is moving both in the forward and
backward directions. Use of resonator gives a great advantage: even several
non-center-symmetrical elementary cells (Fig.\ref{F2}) placed in it can provide the
effect equivalent to that provided by the full-length diffraction grating
(Fig.\ref{F3},\ref{F4}).

For the first view the re-reflection of the wave in resonator (or light
multiple passing over circle resonator of a laser gyroscope) can not provide
the significant increase of the photon path length $L$\ in comparison with
the absorption length $L_{a}$\ because of the absorption of photons in a
medium. Nevertheless this difficulty can be overcome when the part of
resonator is filled by the amplifying medium (for example, inverse medium).
As a result, the electromagnetic wave being absorbed by the investigated gas
is coherently amplified in the amplifier and then is refracted to the gas
again. Consequently, under the ideal conditions the light pulse can exist in
such resonator-amplifier for arbitrarily long time. And if, for example, the
polarization plane of the wave rotates around the $\overrightarrow{E}$\ \
direction, the peculiar ''photon trap'' appears (phase difference of
waves with right (left) circular polarization moving in the opposite
directions in a laser gyroscope or phase difference of waves with \
orthogonal plane polarizations for birefringence effect increases in time).
The angle of rotation $\vartheta ^{T}_{t}=\Omega ^{T}\cdot t$, where $\Omega
^{T} $\ is the frequency of the photon polarization plane rotation around
the $\overrightarrow{E}$\ \ direction, $t$\ is the time of electromagnetic
wave being in a ''trap''. It is easy to find the frequency $\Omega ^{T}$\
from (6): $\ \Omega ^{T}=\frac{\vartheta ^{T}}{L}c=2\pi \rho \omega \beta
_{E}^{T}$. From the estimates of $\vartheta ^{T}$\ it is evident that for $%
\vartheta ^{T}\sim 10^{-12}\;rad$\ (Lead, Yb) \ the frequency $\Omega ^{T}$\
appears to be \ $\Omega ^{T}=\frac{\vartheta ^{T}}{L_{a}}c\sim 10^{-4}\sec
^{-1}$. Therefore $\vartheta ^{T}_{t}\sim 10^{-4}t$\ and for the time $t$ of
about 3 hours the angle $\vartheta ^{T}_{t}$\ becomes $\sim 1\;rad$. The similar
estimates for the atoms Cs, Tl ($\vartheta ^{T}\sim 10^{-13}\;rad$) give
that for the same time the angle $\vartheta ^{T}_{t}\sim 10^{-1}\;rad$.

The time $t$\ is limited, in particular, by spontaneous radiation of photons
in an amplifier that gradually leads to the depolarization of photon gas in
resonator. Surely, it is the ideal picture, but here is the way to further
increase of the experiment sensitivity. The achieved sensitivity in
measurements of phase incursion in laser gyroscope 
makes possible to observe the effect in laser gyroscope, too.
Laser interferometers used as gravitational wave detectors also can provide neccessary sensitivity.

Requiring to measure rotation angle $\sim $10$^{-6}{}rad$\ in "photon trap" and taking into
consideration that existing technique allows to measure much less angles one
can expect to observe effect of the order $\frac{V_{T}}{V_{P}}\sim
10^{-9}\div 10^{-10}.$

Let us consider this question from other point of view. Suppose effects of
polarization plane rotation and birefringence be caused only by atom EDM one
can estimate the possible sensitivity of EDM measurement in such
experiments. Suppose we will
measure rotation angle
 with sensitivity about 10$^{-6}{}rad/hour.$\ Phase difference is $%
\delta \varphi =k(n_{_{\parallel }}-n_{_{\perp }})L=k(n_{_{\parallel
}}-n_{_{\perp }})cT$, where $T$ is the observation time (here $T$ is
supposed to be $T=1\;hour$). Representing $\delta \varphi $\ in the form \ \ 
\begin{equation}
\delta \varphi =\frac{\rho cT\lambda ^{2}}{2\pi }\frac{\Gamma _{e}d_{a}E}{%
\hbar \Gamma ^{2}}  \label{delta fi for EDM}
\end{equation}
where $\rho $\ is the atoms density,\ $\Gamma _{e}$\ is the level radiation
width, $\Gamma $\ is the atom level width (including Doppler widening), \ $E$%
\ is the electric field strength one can estimate $d_{a}$\ as
\begin{equation}
d_{a}=\frac{2\pi \hbar \Gamma ^{2}}{\rho cT\lambda ^{2}\Gamma _{e}E}\delta
\varphi \thickapprox 10^{-33}e  \label{EDM estimation}
\end{equation}
(here $\lambda \sim 10^{-4}cm$, $E=10^{2} CGSE$, $\rho =10^{17}$ $%
atoms/cm^{3}$).

For comparison it is interesting to note that the best expected EDM
measurement limit in recent publications \cite{Budker} is about $\
d_{a}\thickapprox 10^{-28}e$\ , so the advantages of the proposed method
becomes evident.

All the said can be applied not only for the optical range but for the radio
frequency range as well where the observation of the mentioned phenomenon is
also possible by the use of the mentioned atoms and molecules \cite{6}.

Thus, we have shown that the T-odd and P-odd phenomena of photon
polarization plane rotation and circular dichroism in an electric field are
expected to be observable experimentally. The similar situation appears at
the use of non-center-symmetrical diffraction grating.

It should be noted, that the new T-odd and P-odd phenomenon of photon
polarization plane rotation (circular dichroism) in an electric field has
general meaning. Due to quantum electrodynamic effects of electron-positron
pair creation in strong electric, magnetic or gravitational fields, the
vacuum is described by the dielectric permittivity tensor $\varepsilon _{ik}$
depending on these fields\cite{9,13}. The theory of \ $\varepsilon _{ik}$ 
\cite{9,13} does not take into account the weak interaction of electron and
positron with each other. Considering the weak interaction between electron
and positron in the process of pair creation in an electric (gravitational)
field one can obtain that the permittivity tensor of vacuum in strong
electric (gravitational) field contains the term $\varepsilon
_{ik}^{vac}\sim i\beta _{vac\overrightarrow{E}}^{T}\varepsilon _{ikl}n_{lE}$
($\varepsilon _{ik}^{vac}\sim i\beta _{vac\overrightarrow{g}}^{T}\varepsilon
_{ikl}n_{lg},\overrightarrow{n_{g}}=\frac{\overrightarrow{g}}{g},%
\overrightarrow{g}$ is the free fall acceleration), and as a result, the
polarization plane rotation (circular dichroism) phenomena exist for photons
moving in an electric (gravitational) field in vacuum. And visa versa\ $%
\gamma $-quanta appearing under single-photon electron-positron annihilation
in an electric (gravitational) field have the admixture of circular
polarization, caused by T-odd P-odd weak interactions.

\newpage

{\bf {\Large 4. Phenomenon of the time-reversal violating magnetic field
generation by a static electric field in a medium and vacuum.}}

\bigskip

As it was shown P- and T-odd interactions cause mixing of opposite parity levels of atom
(molecule) that yields to the appearance of P- and T-odd terms of the atom
(molecule) polarizability \cite{4}. This makes possible to observe various
optical phenomena, for example, photon polarization plane rotation (birefringence and
circular dichroism) in an optically homogeneous medium placed to an electric
field.

The energy of atom (molecule) in external electromagnetic field includes the
term caused by the time reversal violating interactions \cite{4}:
\begin{equation}
\Delta U=-\frac{1}{2}\beta _{S}^{T}\overrightarrow{E}\overrightarrow{H},
\label{1}
\end{equation}
where $\beta _{S}^{T}$ is the scalar T-noninvariant polarizability of atom
(molecule), $\overrightarrow{E}$\ \ is the external electric field, $%
\overrightarrow{H}$\ is the external magnetic field.

It's well known \cite{8} that when the external field frequency $\omega
\rightarrow 0$ the polarizabilities describe the processes of magnetization
of medium by a static magnetic field and electric polarization of a medium
by a static electric field

The energy of interaction of magnetic moment $\overrightarrow{\mu }$ with
magnetic field $\overrightarrow{H}$
\begin{equation}
W_{H}=-\overrightarrow{\mu }\overrightarrow{H}  \label{2}
\end{equation}
Comparison of (\ref{1}) and (\ref{2}) let one to conclude that the action of
stationary electric field on an atom (molecule) induces the magnetic moment
of atom
\begin{equation}
\overrightarrow{\mu_{ind}}(\overrightarrow{E})=\frac{1}{2}\beta _{S}^{T} 
\overrightarrow{E}
\end{equation}
On the other hand, the energy of interaction of electric dipole moment $%
\overrightarrow{d}$ with electric field $\overrightarrow{E}$
\begin{equation}
W_{E}=-\overrightarrow{d}\overrightarrow{E}.  \label{4}
\end{equation}
As it follows from (\ref{1}) and (\ref{4}), magnetic field induces the
electric dipole moment of atom
\begin{equation}
\overrightarrow{d_{ind}}(\overrightarrow{H})=\frac{1}{2}\beta _{S}^{T}%
\overrightarrow{H}
\end{equation}
As appears from the above, atom (molecule) being placed to static electric
field gets the induced magnetic moment which in its part produces magnetic
field. And similarly, if atom (molecule) is exposed to magnetic field the
induced electric dipole moment yields to the appearance of its associated
electric field.

Let us consider the simplest possible experiment. Suppose that homogeneous
isotropic matter (liquid or gas) is exposed to an electric field $%
\overrightarrow{E}$. From the above it follows that the time reversal
violation yields to the appearance of magnetic field $\overrightarrow{H}%
_{T}=4\pi \rho \overrightarrow{\mu }(\overrightarrow{E})$ parallel to $%
\overrightarrow{E}$\ in this area ($\rho $ is the number of atoms
(molecules) of matter per $cm^{3}$ ). And vice versa, the electric field $%
\overrightarrow{E}_{T}=4\pi \rho \overrightarrow{d}(\overrightarrow{H})$
appears under matter placement to the area occupied by a magnetic field $%
\overrightarrow{H}$. Let us estimate the effect value. It is easy to do by $%
\beta _{S}^{T}$ evaluation. The general case explicit expression for
polarizabilities for time dependent fields were derived in \cite{4} (see
eqns. (12)-(20) therein).

Briefly the calculation technique is as follows. Let us suppose that atom is
placed to the arbitrary periodic in time electric and magnetic fields. The
energy of interaction of an atom (molecule) with these fields has the
routine form
\begin{equation}
W=-\widehat{\overrightarrow{d}}\overrightarrow{E}-\widehat{\overrightarrow{
\mu }}\overrightarrow{H}+.....  \label{interaction energy}
\end{equation}
where $\widehat{\overrightarrow{d}}$ is the operator of atom electric dipole
moment and $\widehat{\overrightarrow{\mu }}$ is the operator of atom
magnetic dipole moment 
\begin{equation}
\overrightarrow{E}=\frac{1}{2}\left\{ \overrightarrow{E}_{0}\;e^{-i\omega
t}+ \overrightarrow{E}_{0}^{\ast }\;e^{i\omega t}\right\} ,\;\overrightarrow{
H}= \frac{1}{2}\left\{ \overrightarrow{H}_{0}\;e^{-i\omega t}+ 
\overrightarrow{H} _{0}^{\ast }\;e^{i\omega t}\right\}
\end{equation}
The Shr\"{o}dinger equation describing atom interaction with electromagnetic
field is as follows:
\begin{equation}
i\hbar \frac{\partial \psi (\xi ,t)}{\partial t}=[H_{A}(\xi )+W(\xi ,t)]\psi
(\xi ,t),
\end{equation}
where $H_{A}(\xi )$ is the atom Hamiltonian taking into account the weak
interaction of electrons with nucleus in the center of mass of the system, $%
\xi $ is the space and spin variable of electron and nucleus, $W$ is the
energy of interaction of atom with electromagnetic field of frequency $%
\omega $
\begin{eqnarray}
W &=&Ve^{-i\omega t}+V^{+}e^{i\omega t},  \label{W} \\
V &=&-\frac{1}{2}(\overrightarrow{d}\overrightarrow{E_{0}}+\overrightarrow{%
\mu }\overrightarrow{H_{0}}),V^{+}=-\frac{1}{2}(\overrightarrow{d}%
\overrightarrow{E_{0}}^{\ast }+\overrightarrow{\mu }\overrightarrow{H_{0}}%
^{\ast })  \nonumber
\end{eqnarray}
Let us perform the transformation $\psi =\exp (-\frac{{iH_{A}t}}{{\hbar }}%
)\varphi $. Suppose $H_{A}\psi _{n}=E_{n}\psi _{n}$ ($E_{n}=E_{n}^{(0)}-%
\frac{1}{2}i\Gamma _{n}$, $E_{n}^{(0)}$ is the atom level energy, $\Gamma
_{n}$ is the atom level width), then $\varphi =\sum_{n}b_{n}(t)\psi _{n}$.
Therefore it follows from (\ref{W}) 
\begin{eqnarray}
i\hbar \frac{\partial b_{n}(t)}{\partial t}=\sum_{f}\left\{ \langle
n|V|f\rangle \exp [i(E_{n}-E_{f}-\hbar \omega )t/\hbar ]\right. + \\
+\left. \langle n|V^{+}|f\rangle \exp [i(E_{n}-E_{f}+\hbar \omega )t/\hbar
]\right\} b_{f}(t),\;\langle \psi _{n}|\psi _{m}\rangle \ll 1.  \nonumber
\end{eqnarray}
Suppose $b_{n0}$ be the ground state amplitude. Let us substitute the
amplitude $b_{f}$ describing the excited atom state into the equation for $%
b_{n0}$ and study this equation\ at time $t\gg \tau _{f}=\hbar /\Gamma
_{f}\;($or $\tau _{f}=\hbar /\Delta E);\;\Delta E=E_{f}^{(0)}-E_{n0}-\hbar
\omega $; $\Gamma _{f}$ $\gg $ $|\langle n|V|f\rangle |$ (or $\Delta E\gg $ $%
|\langle n|V|f\rangle |$). Therefore $b_{n0}$ is defined by equation 
\[
i\hbar \frac{\partial b_{n0}(t)}{\partial t}=\widehat{U}_{eff}\;b_{n0},{\
where} 
\]
\begin{equation}
\widehat{U}_{eff}=-\sum_{f}\left( \frac{\langle n_{0}|V|f\rangle \;\langle
f|V^{+}|n_{0}\rangle }{E_{f}-E_{n0}+\hbar \omega }+\frac{\langle
n_{0}|V^{+}|f\rangle \;\langle f|V|n_{0}\rangle }{E_{f}-E_{n0}-\hbar \omega }%
\right)  \label{U_eff}
\end{equation}
Substituting $V$ and $V^{+}$ into\ (\ref{U_eff}) one can obtain 
\begin{equation}
\widehat{U}_{eff}=-\frac{1}{2}\widehat{g}_{ik}^{E}E_{0i}E_{0k}^{\ast }-\frac{%
1}{2}\widehat{g}_{ik}^{H}H_{0i}H_{0k}^{\ast }-\frac{1}{2}\widehat{g}%
_{ik}^{EH}E_{0i}H_{0k}^{\ast }-\frac{1}{2}\widehat{g}%
_{ik}^{HE}H_{0i}E_{0k}^{\ast },
\end{equation}
where the polarizability of atom (molecule) is: 
\begin{eqnarray}
\widehat{g}_{ik}^{E} &=&-\frac{1}{2}\left( \sum_{f}\frac{\langle
n_{0}|d_{i}|f\rangle \;\langle f|d_{k}|n_{0}\rangle }{E_{f}-E_{n0}+\hbar
\omega }+\frac{\langle n_{0}|d_{k}|f\rangle \;\langle f|d_{i}|n_{0}\rangle }{%
E_{f}-E_{n0}-\hbar \omega }\right)  \nonumber \\
\widehat{g}_{ik}^{H} &=&-\frac{1}{2}\left( \sum_{f}\frac{\langle n_{0}|\mu
_{i}|f\rangle \;\langle f|\mu _{k}|n_{0}\rangle }{E_{f}-E_{n0}+\hbar \omega }%
+\frac{\langle n_{0}|\mu _{k}|f\rangle \;\langle f|\mu _{i}|n_{0}\rangle }{%
E_{f}-E_{n0}-\hbar \omega }\right)  \nonumber \\
\widehat{g}_{ik}^{EH} &=&-\frac{1}{2}\left( \sum_{f}\frac{\langle
n_{0}|d_{i}|f\rangle \;\langle f|\mu _{k}|n_{0}\rangle }{E_{f}-E_{n0}+\hbar
\omega }+\frac{\langle n_{0}|\mu _{k}|f\rangle \;\langle
f|d_{i}|n_{0}\rangle }{E_{f}-E_{n0}-\hbar \omega }\right)  \nonumber \\
\widehat{g}_{ik}^{HE} &=&-\frac{1}{2}\left( \sum_{f}\frac{\langle n_{0}|\mu
_{i}|f\rangle \;\langle f|d_{k}|n_{0}\rangle }{E_{f}-E_{n0}+\hbar \omega }+%
\frac{\langle n_{0}|d_{k}|f\rangle \;\langle f|\mu _{i}|n_{0}\rangle }{%
E_{f}-E_{n0}-\hbar \omega }\right)  \nonumber
\end{eqnarray}
It should be noted that $\widehat{g}_{ik}^{E}$ and $\widehat{g}_{ik}^{H}$
are the P- and T-invariant electric and magnetic polarizability tensors and $%
\widehat{g}_{ik}^{EH}$ and $\widehat{g}_{ik}^{HE}$ are the P- and
T-noninvariant polarizability tensors

Let an atom be placed at the static ($\omega \rightarrow 0$) magnetic and
electric fields $\overrightarrow{E}$ and $\overrightarrow{H}$ of the same
direction. Then it's perfectly easy to obtain the effective energy of P- and
T-odd interaction of an atom with these fields.
\begin{equation}
\widehat{U}_{eff}^{T,P}=-\frac{1}{2}\left( \sum_{f}\frac{\langle
n_{0}|d_{z}|f\rangle \;\langle f|\mu _{z}|n_{0}\rangle +\langle n_{0}|\mu
_{z}|f\rangle \;\langle f|d_{z}|n_{0}\rangle }{E_{f}-E_{n_{0}}}\right) EH
\end{equation}
Axis $z$ is supposed to be parallel to $\overrightarrow{E}$. Thus from (\ref
{1}) 
\begin{equation}
\beta _{S}^{T}=\sum_{f}\frac{\langle n_{0}|d_{z}|f\rangle \;\langle f|\mu
_{z}|n_{0}\rangle +\langle n_{0}|\mu _{z}|f\rangle \;\langle
f|d_{z}|n_{0}\rangle }{E_{f}-E_{n_{0}}}
\end{equation}
Let us estimate the $\beta _{S}^{T}$ order of magnitude. The atom state \ $%
|f\rangle $\ does not possess the certain parity because of weak T-odd
interactions. And over the weakness of $V_{T}$\ the state $|f\rangle $\ is
mixed with the opposite parity state by factor of $\eta _{T}=\frac{{V_{W}^{T}}%
}{{E_{f}-E_{n}}}$. According to the above 
\begin{equation}
\beta _{S}^{T}\sim \frac{\langle d\rangle \;\langle \mu \rangle }{%
E_{f}-E_{n_{0}}}\eta _{T}
\end{equation}

For the heavy atoms the mixing coefficient can attain the value $\eta
_{T}\approx 10^{-14}$. Taking into account that matrix element $\langle \mu
\rangle \sim \alpha \langle d\rangle $ (where $\alpha =\frac{1}{137}$ is the
fine structure constant) one can obtain $\beta _{S}^{T}\sim \eta
_{T}\;\alpha \frac{{\langle d\rangle ^{2}}}{{\Delta }}{\;}\approx
10^{-16}\cdot \frac{{8\cdot 10^{-36}}}{{10^{-12}}}\approx 10^{-40}$.
Therefore, the electric field $E=10^{2}\;CGSE$ induces magnetic moment $\mu
_{T}\approx 10^{-38}$. Then, the magnetic field in the liquid target can be
estimated as follows
\begin{equation}
H=4\pi \rho \mu _{T}\approx 10^{23}\cdot 10^{-38}=10^{-15}\;gauss
\end{equation}
The magnitude of magnetic field strength can be increased, for example, by
tightening of the magnetic field with superconductive shield. In this way
the measured field strength can be increased by four orders when one collect
the field from the area 1 $m^{2}$ to the area 1 $cm^{2}$ (Fig.\ref{5}).

\begin{figure}[h]
\epsfxsize = 10 cm \centerline{\epsfbox{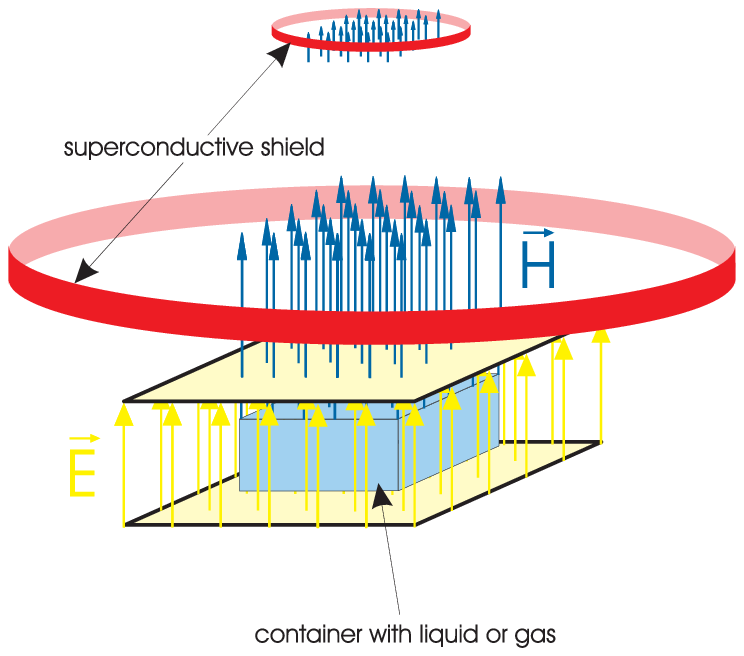}}
\caption{ }
\label{F5}
\end{figure}

The induced magnetic moment produces magnetic field at the electron
(nucleus) of the atom. This field $H^{T}(E)\sim \mu \;\langle \frac{{1}}{{%
r^{3}}}\rangle \sim 10^{-38}\cdot 10^{26}=10^{-12}$ $gauss$. Therefore, the
frequency of precession of atom magnetic moment $\mu _{A}$ in the magnetic
field induced by an external electric field
\begin{equation}
\Omega _{E}\sim \frac{\mu _{A}\;\beta \;E\;\langle {1/}{r^{3}}\rangle }{%
\hbar }=\frac{10^{-20}\cdot 10^{-12}}{10^{-27}}=10^{-5}\;\sec ^{-1}
\end{equation}
It should be reminded that to measure the electric dipole moment the shift
of precession frequency of atom spin in the presence of both magnetic and
electric fields is investigated. Then, the T-odd shift of precession
frequency of atom spin includes two terms: frequency shift conditioned by
interaction of atom electric dipole moment with electric field $\omega
_{E}\sim {d_{A}E/}{\hbar }$ and frequency shift $\Omega \sim {\mu H^{T}(E)}{%
\ /\hbar }$ defined above. This aspect should be considered when
interpreting the similar experiments. One should take note of the mixing
coefficient $\eta _{T}$ essential increase when the opposite parity levels
are closed to each other or even degenerate. Then the effect can grow up as
much as several orders $10^{5}\div 10^{6}$ (this occurs, for example, for
Dy, TlF, BiS, HgF).

The similar phenomenon of magnetic field induction by electric field can
occur in vacuum too.

Due to quantum electrodynamic effect of electron-positron pair creation in
strong electric, magnetic or gravitational field, the vacuum is described by
the dielectric $\varepsilon _{ik}$ and magnetic $\mu _{ik}$\ permittivity
tensors depending on these fields. The theory of $\varepsilon _{ik}$ \cite{9}
does not take into account the weak interaction of electron and positron
with each other. Considering the T- and P-odd weak interaction between
electron and positron in the process of pair creation in an electric
(magnetic, gravitational) field one can obtain the density of
electromagnetic energy of vacuum contains term $\beta _{v}^{T}(%
\overrightarrow{E}\overrightarrow{H)}$ similar (\ref{1}) (in the case of
vacuum polarization by a stationary gravitational field $\beta _{g}^{T}(%
\overrightarrow{H}\overrightarrow{n_{g}})$, $\overrightarrow{n_{g}}=\frac{%
\overrightarrow{g}}{g}$, $\overrightarrow{g}-$ gravitational acceleration).

As a result both electric and magnetic fields (directed along the electric
field) could exist around an electric charge. 
But in this case $\oint \overrightarrow{B}d\overrightarrow{S}\neq 0$ ($%
\overrightarrow{B}$\ is the magnetic induction) that is impossible in the
framework of classic electrodynamics. The existence of such field would
means the existence of induced magnetic monopole. If the condition $\oint 
\overrightarrow{B}d\overrightarrow{S}=0$ is fulfilled then for the
spherically symmetrical case the field appears equal to zero. Surely, the
value of this magnetic field is extremely small, but the possibility of its
existence is remarkable itself.

The above result can be obtained in the framework of general Lagrangian
formalism. Lagrangian density can depend only on the field invariants. Two
invariants are known for the quasistatic electromagnetic field: $(%
\overrightarrow{E}\overrightarrow{H})$ and $(E^{2}-H^{2})$. In conventional
T-invariant theory these invariants are included in the Lagrangian $L$ only
as $(E^{2}-H^{2})$ and $(\overrightarrow{E}\overrightarrow{H})^{2}$, i.e. $%
L=L(E^{2}-H^{2},(\overrightarrow{E}\overrightarrow{H})^{2})$ \cite{9}. But
while taking into account the T-odd interactions the Lagrangian can include
\ invariant $(\overrightarrow{E}\overrightarrow{H})$ raising to the odd
power, i.e.
\begin{equation}
L_{T}=L_{T}(E^{2}-H^{2},(\overrightarrow{E}\overrightarrow{H})^{2},( 
\overrightarrow{E}\overrightarrow{H}))  \label{lagrangian}
\end{equation}
Expanding (\ref{lagrangian}) by weak interaction one can obtain
\begin{equation}
L_{T}=L(E^{2}-H^{2},(\overrightarrow{E}\overrightarrow{H})^{2})+\beta _{T}(%
\overrightarrow{E}\overrightarrow{H}),
\end{equation}
where $L$ is the density of Lagrangian in P- and T-invariant
electrodynamics, $\beta _{T}=\beta _{T}(E^{2}-H^{2},(\overrightarrow{E}%
\overrightarrow{H})^{2})$ \ is the constant \ can be found in certain
theory. The explicit form of $L$ is cited in \cite{9}.

The additions caused by the vacuum polarization can be described by the
field dependent dielectric and magnetic permittivity of vacuum. According to 
\cite{9} the electric induction vector $\overrightarrow{D}$ and magnetic
induction vector $\overrightarrow{B}$ are defined as:
\begin{equation}
\overrightarrow{D}=\frac{\partial L}{\partial \overrightarrow{E}},\;%
\overrightarrow{B}=-\frac{\partial L}{\partial \overrightarrow{H}}
\end{equation}
Similarly the electric polarization $\overrightarrow{P}$ and magnetization $%
\overrightarrow{M}$ of vacuum can be found \cite{9}:
\begin{eqnarray}
\overrightarrow{P} &=&\frac{\partial (L_{T}-L_{0})}{\partial \overrightarrow{%
E}},\;\overrightarrow{M}=-\frac{\partial (L_{T}-L_{0})}{\partial 
\overrightarrow{H}}, \\
\overrightarrow{D} &=&\overrightarrow{E}+4\pi \overrightarrow{P},\;%
\overrightarrow{B}=\overrightarrow{H}+4\pi \overrightarrow{M.}
\end{eqnarray}
In accordance with the above, the T-noninvariance yields to the appearance
of additional P- and T-odd terms to the electric polarization $%
\overrightarrow{P}$ and magnetization $\overrightarrow{M}$ . There are the
addition to the vector of electric polarization $\overrightarrow{P}$
proportional to the magnetic field strength $\overrightarrow{H}$ and the
addition to the vector of magnetization $\overrightarrow{M}$ proportional to
the electric field strength $\overrightarrow{E}$

\end{document}